\documentclass{iopart}
\usepackage{geometry}
\usepackage{graphicx, bm, amssymb, xcolor}
\usepackage{verbatim}
\usepackage[breaklinks,colorlinks,citecolor=blue]{hyperref}
\usepackage[all]{hypcap}

\usepackage{fancyhdr}
\usepackage{aas_macros}
\usepackage{epsfig}
\usepackage{natbib}
\usepackage{framed}
\usepackage{times}
\usepackage{iopams}

\newcommand{\msun}{\mbox{M$_\odot$}}
\newcommand{\yr}{\mbox{${\rm yr}$}}
\newcommand{\myr}{\mbox{${\rm Myr}$}}
\newcommand{\gyr}{\mbox{${\rm Gyr}$}}
\newcommand{\pc}{\mbox{${\rm pc}$}}
\newcommand{\kpc}{\mbox{${\rm kpc}$}}
\newcommand{\kms}{\mbox{${\rm km}~{\rm s}^{-1}$}}
\newcommand{\cmc}{\mbox{${\rm cm}^{-3}$}}
\newcommand{\feh}{\mbox{$[{\rm Fe}/{\rm H}]$}}
\newcommand{\oh}{\mbox{$[{\rm O}/{\rm H}]$}}
\newcommand{\zh}{\mbox{$[{\rm Z}/{\rm H}]$}}
\newcommand{\afe}{\mbox{$[\alpha/{\rm Fe}]$}}

\begin{document}

\paper[Globular cluster formation]{Globular cluster formation in the context of galaxy formation and evolution}
\author{J.~M.~Diederik Kruijssen}
\address{Max-Planck Institut f\"ur Astrophysik, Garching, Germany}
\ead{kruijssen@mpa-garching.mpg.de}

\begin{abstract}
The formation of globular clusters (GCs) remains one of the main unsolved problems in star and galaxy formation. The past decades have seen important progress in constraining the physics of GC formation from a variety of directions. In this review, we discuss the latest constraints obtained from studies of present-day GC populations, the formation of young massive clusters (YMCs) in the local Universe, and the observed, large-scale conditions for star and cluster formation in high-redshift galaxies. The main conclusion is that the formation of massive, GC progenitor clusters is restricted to high-pressure environments similar to those observed at high redshift and at the sites of YMC formation in the local Universe. However, the correspondingly high gas densities also lead to efficient cluster disruption by impulsive tidal shocks, which limits the survival of GCs progenitor clusters. As a result, the long-term survival of GC progenitor clusters requires them to migrate into the host galaxy halo on a short time-scale. It is proposed that the necessary cluster migration is facilitated by the frequent galaxy mergers occurring at high redshift. We use the available observational and theoretical constraints to condense the current state of the field into a coherent picture of GC formation, in which regular star and cluster formation in high-redshift galaxies naturally leads to the GC populations observed today.
\end{abstract}


\section{Introduction} \label{sec:intro}
In 1789, William Herschel presented his second catalogue of 1000 `nebulae' \citep{herschel89}, in which he reported that dozens of spherical examples could be resolved into collections of thousands of stars. He named these objects `globular clusters' (GCs)\footnote{In the same treatise, Herschel also for the first time coined the term `planetary nebula'.} and suggested that their centrally condensed morphology was the result of a central, attractive force, which would lead them to become more concentrated (and spherical) with age. Centuries later, Herschel's ideas hold up remarkably well, although our understanding of the dynamical evolution of GCs has advanced greatly since his early physical analysis \citep[see e.g.][]{heggie03}. The past decades in particular have seen a flurry of work on the present-day properties of GC populations and their evolution within the dark matter haloes of galaxies. However, the physics of GC formation remains a mystery.

Motivated by the properties of the nearest dense stellar systems in the Galactic halo, the term `globular cluster' has traditionally been used according to a wide variety of definitions, e.g.~based on metallicity (`metal-poor' or `sub-solar'), mass ($M=10^4$--$10^6~\msun$), age ($\tau\sim10^{10}~\yr$), location (`in the halo'), or chemical composition (`Na-O anti-correlation'), although exceptions to these criteria are known \citep[e.g.][]{harris96,dinescu99,forbes10}. Modern observational facilities cannot resolve the GC formation process -- given the old ages typically assigned to GCs, they must have formed at redshifts $z>2$. At these redshifts, the present-day sizes of GCs and giant molecular clouds (GMCs) of $10$--$100~\pc$ correspond to angular sizes of $10^{-3}$--$10^{-2}$~arcseconds. The unresolved nature of GC formation requires that the relevant physical processes are inferred indirectly. In this review, the current constraints on the physics of GC formation will be discussed from three directions: the formation of massive clusters in the local Universe, the observed conditions for star and cluster formation in high-redshift galaxies, and the present-day properties of GC populations. Throughout, it will be reiterated that GC formation is fundamentally a two-step process: (1) enough mass must be accumulated to form a massive stellar cluster, and (2) the cluster must survive for almost a Hubble time to be observable at the present day.

Originally, GCs were thought to be unique relics of early star and galaxy formation, which could only form due to the special conditions at high-redshift, such as the high Jeans mass following recombination \citep{peebles68} or the presence of thermally unstable, metal-poor gas in galactic haloes \citep{fall85}. However, the 1980s and especially the 1990s witnessed the discovery of young massive clusters (YMCs; masses $M=10^4$--$10^8~\msun$ and ages $\tau<1~\gyr$) in galaxy mergers and merger remnants throughout the local Universe \citep[e.g.][]{schweizer82,schweizer87,holtzman92,schweizer96,whitmore99,bastian06}, suggesting that GC-like stellar clusters could still form in the Universe today. For the first time, the formation of clusters with masses similar to (or exceeding) those of GCs were observed, but only in galaxies with high gas densities, turbulent velocities, and hence gas pressures. These conditions were common in star-forming galaxies at the peak of the cosmic star formation history at $z=2$--$3$ \citep[e.g.][]{hopkins06b,tacconi08,daddi10b,tacconi10,genzel10,swinbank11}. Occam's Razor (`plurality must never be posited without necessity') therefore suggests that the primary question to ask is: {\it could regular YMC formation physics in $z>2$ galaxies have led to the GC populations residing in galaxies today?}

\begin{figure}
\center\includegraphics[height=12cm,angle=90]{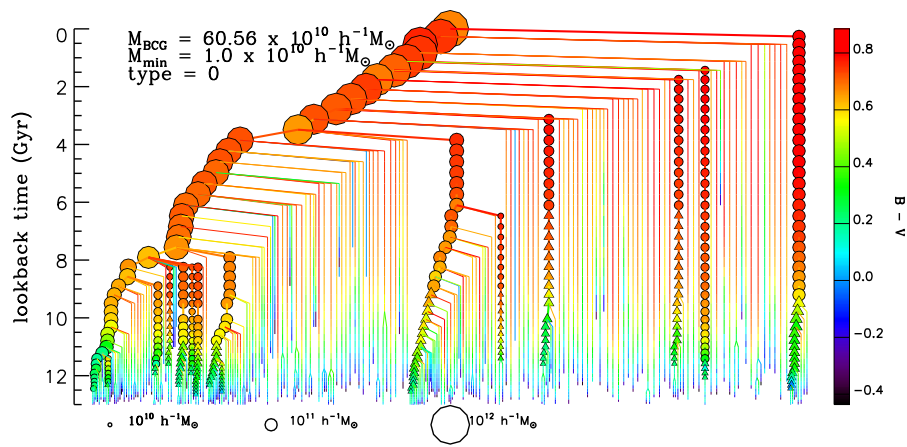}
\caption{This figure shows the galaxy merger tree of a brightest-cluster galaxy from the Millennium Simulation \citep{springel05d}, illustrating that galaxies grow by the hierarchical merging with other galaxies, which occurs particularly often at early cosmic times. These galaxy mergers are the likely agents for the redistribution of massive stellar clusters into the haloes of their host galaxies, where GC populations presently reside. Lines and symbols indicate different galaxies, colour-coded by their $B-V$ colour and scaled by their stellar mass as indicated by the legends (the information on the baryonic content is obtained with a semi-analytic model). Symbols are omitted for haloes less massive than $10^{10}~h^{-1}~\msun$. Taken from \citet[Figure~1]{delucia07}, reproduced with permission.} \label{fig:tree}
\end{figure}

There have been two recent major reviews on the formation \citep{longmore14} and evolution \citep{portegieszwart10} of YMCs. Stimulated by the discovery of these `young GCs' and the two decades of YMC studies following the discovery, both reviews advocate the interpretation that GCs are indeed old YMCs, although no quantitative evidence is provided beyond the quantitative similarities in their masses, sizes, and stellar populations.\footnote{Young {\it low-mass} clusters are excluded from the analogy because of their short lifetimes. For instance, clusters with masses $M<10^4~\msun$ survive for less than $1~\gyr$ in the solar neighbourhood \citep{lamers06a}, whereas even shorter lifetimes are reached in higher-density environments \citep{gieles06,gieles07,kruijssen11}. It is therefore unlikely that even the lowest-mass GCs observed today had initial masses $M_{\rm init}<10^4~\msun$.} Turning this (well-informed) conjecture into a physical model of GC formation requires an answer to a fundamental question: how did YMCs become GCs? The present-day spatial distribution of GCs in the haloes of nearby galaxies suggests that (Y)MCs and the star-forming components of their host galaxies must have become displaced. Given the frequent galaxy mergers and the resulting morphological transformations that characterise galaxy evolution in a $\Lambda$CDM Universe \citep[see Figure~\ref{fig:tree} and e.g.][]{sales10,vogelsberger14}, it is inevitable that there is a reasonable degree of migration of (Y)MCs away from a galaxy's star-forming component. As we shall see in \S\ref{sec:pic}, this migration could play a central role in the long-term survival of GCs.

Of course, any constraints put on GC formation by considering YMC formation in $z>2$ galaxies must be evaluated using present-day GC populations. This requires a description of GC survival and destruction over almost a Hubble time of dynamical evolution. Whereas the evolution of GCs in present-day haloes has been studied in considerable detail, the early evolution of (Y)MCs in the high-pressure $z>2$ environment is still an emerging field. The distinction is important, because the high pressures ($P/k\sim10^7~{\rm K}~\cmc$ as opposed to $P/k\sim10^4$--$10^5~{\rm K}~\cmc$ in the Milky Way) and densities ($\Sigma\sim10^2$--$10^{3.5}~\msun~\pc^{-2}$ as opposed to $\Sigma\sim10~\msun~\pc^{-2}$ in the Milky Way) seen in $z>2$ environments are more disruptive to stellar clusters than the haloes in which they presently reside \citep{elmegreen10,kruijssen12c}. While the violent conditions may have similarly enabled the formation of GC progenitors as they promote YMC formation in the nearby Universe, the environment likely also destroyed some fraction of the young GC population. It is therefore important to quantify how much time was spent by (Y)MCs in high-density environments, what these environments looked like, and how the GCs subsequently evolved until the present day.

The physics of GC formation is also important for studies of the abundant exotica hosted by GCs, such as blue stragglers \citep[e.g.][]{ferraro14}, (single or binary) pulsars \citep[e.g.][]{davies98,pfahl02} and stellar mass black holes \citep[e.g.][]{moody09,strader12}, or theoretically predicted objects such as intermediate-mass black holes \citep{portegieszwart04,luetzgendorf13}. The violent internal dynamics of GCs provide efficient formation channels for these compact objects in general and compact binaries in particular, which can be used to address fundamental questions such as the nature of gravity in the extremely relativistic regime, highlighting GCs as unique cosmic laboratories \citep[e.g.][]{portegieszwart10}. An obvious implication is that the cosmic distribution of relativistically interacting cosmic objects is affected by the formation and evolution of the GC population. If GCs at high redshift formed and evolved in the same way as YMCs in the local Universe, this may present an opportunity to understand and quantify the formation and distribution of the compact binary population. For instance, it is well-known that cluster disruption has shaped the present-day GC population -- when and under which conditions this disruption took place determines how the numbers, masses, and densities of GCs have changed with redshift, and hence sets the cosmic abundance of GC-specific channels for compact binary formation. As a result, obtaining an understanding of the formation, evolution and cosmic (re)distribution of GCs is key for future surveys aiming to detect the emission of gravitational waves by mergers between compact objects.

It is the goal of this work to review and quantify the current observational and theoretical constraints on the physics of GC formation, in order to construct a comprehensive scenario for the formation of GCs. In \S\ref{sec:obs}, we will first discuss the observational picture, starting with the formation of YMCs in the local Universe (\S\ref{sec:ymc}), before turning to the conditions of star formation in high-redshift galaxies (\S\ref{sec:hiz}), and finishing with the properties of present-day GC systems (\S\ref{sec:gc}). The physical mechanisms governing the formation and evolution of massive stellar clusters will be discussed in \S\ref{sec:phys}. In \S\ref{sec:pic}, we will condense the current state of the field into a coherent picture of GC formation.

\section{Observational constraints on globular cluster formation} \label{sec:obs}
\begin{figure}
\center\includegraphics[width=11.4cm]{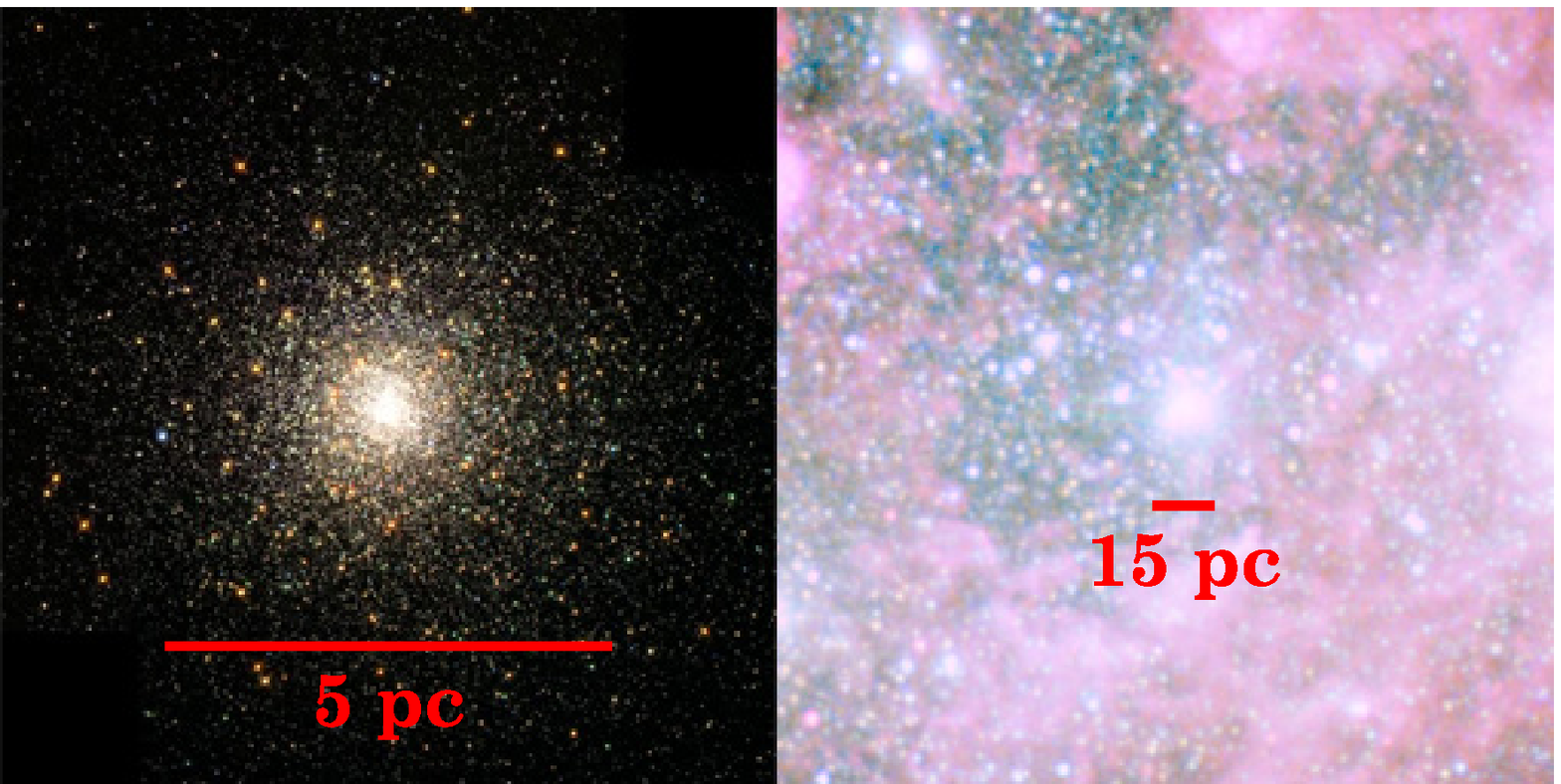}
\caption{This figure illustrates the similarities between the Galactic GC M80 (left) and the most massive YMC in the nearby dwarf starburst galaxy NGC~1569 (right). The masses ($M=\{10^{5.6},10^6\}~\msun$) and half-light radii ($R=\{1.8,1.6\}~\pc$) are very similar \citep{harris96,anders04}, but the ages ($\tau=\{10^{10},10^7\}~\yr$) differ strongly, showing that GC-like YMCs are still forming at the present day. Left image credit: NASA and The Hubble Heritage Team (STScI/AURA). Right image credit: NASA, ESA, the Hubble Heritage Team (STScI/AURA), and A. Aloisi (STScI/ESA).} \label{fig:clusters}
\end{figure}
There are three main avenues to constrain the physics of GC formation observationally.
\begin{itemize}
\item[(i)]
One of the main discoveries of the Hubble Space Telescope (HST) has been that clusters with masses similar to (or higher than) those of GCs are still forming today \citep[see Figure~\ref{fig:clusters} and e.g.][but also see \citealt{schweizer82,schweizer87} for important pre-HST work]{holtzman92,schweizer96,whitmore99,bastian06}, covering a broad range of metallicities that reflects the wide variety of star-forming galaxies in the local Universe. Numerous authors have since suggested that YMCs are the progenitors of future `GC populations' \citep{ashman92,elmegreen97,portegieszwart10,longmore14}. If this analogy holds, the formation of YMCs in the nearby Universe may be used to infer the conditions of GC formation.
\item[(ii)]
The old ages of Local-Group GCs \citep[$\tau=5$--$13~\gyr$,][]{forbes10} imply that they did not form in the low-pressure environments seen in nearby galaxies, but originate from the high-pressure conditions seen at high redshift ($z=2$--$6$), during the peak of the cosmic star formation history \citep{hopkins06b}. State-of-the-art observational facilities like the Atacama Large Millimeter Array (ALMA), the Very Large Telescope (VLT) and HST enable the star formation process in high-redshift galaxies to be probed directly. Both observational \citep[e.g.][]{shapiro10} and theoretical \citep{li04,bournaud08,elmegreen10,kruijssen12c} studies have shown that the conditions in these galaxies can put important constraints on GC formation, such as their possible formation sites within galaxies and the implications thereof for GC survival.
\item[(iii)]
The conditions of GC formation may be inferred from their present-day properties, such as their mass function, specific frequency (number of GCs per unit stellar light of the host galaxy), or chemical abundance patterns \citep{harris02,brodie06,jordan07,gratton12}. This potential avenue has been enabled by systematic surveys of Galactic and extragalactic GC populations using HST, Keck, Gemini, Subaru, and many other observatories. However, tracing the physics of GC formation using present-day GC populations is only possible if {\it either} the observable in question has been unaffected by a Hubble time of GC evolution \citep[the chemical composition may be a good example, although GC evaporation can cause the number ratios of chemical sub-populations of stars to evolve substantially if they are dynamically not well-mixed, see e.g.][]{gratton12} {\it or} a reliable model for the time-evolution of the observable in question allow the initial conditions to be reverse-engineered. Needless to say, this can be a highly complex exercise requiring substantial input from theory, of which it is hard to be certain if all relevant physical processes have been included.
\end{itemize}
Because all three of the above avenues do not probe GC formation directly, it is crucial that the derived constraints on GC formation are mutually consistent. In case of discrepancies, the most direct constraints should take precedence over the others. In practice, this means that avenues~(i) and~(ii) most directly constrain the physical formation conditions of GCs (gas pressure, turbulence, metallicity, host galaxy, etc.), whereas avenue~(iii) puts constraints on the outcome of these processes (ages, masses, chemical abundance patterns, etc.). Inferring the conditions of GC formation from avenue~(iii) may not lead to unique solutions. The logical course of action as a community is therefore to use avenues~(i) and~(ii) to formulate hypotheses for the physics of GC formation, which need to be consistent with the outcome after a Hubble time as characterised through avenue~(iii). In other words: could the products of YMC formation in the high-redshift Universe have survived until the present day?

\subsection{Nearby young massive cluster formation} \label{sec:ymc}
\citet{portegieszwart10} and \citet{longmore14} have recently reviewed the formation and evolution of YMCs at length. Here, we will only highlight those aspects most relevant for GC formation.

\subsubsection{Hierarchical initial conditions} \label{sec:hierarchy}
On spatial scales smaller than the scale height of a galactic disc, the interstellar medium (ISM) is structured hierarchically. Observationally, YMCs are thought to form in a similar hierarchy, through the merging of smaller stellar groups \citep{elmegreen96,efremov98,hopkins13}. ALMA observations of a dense, proto-YMC cloud have now confirmed that the hierarchy continues down to sub-$\pc$ scales even in the highest-density environments \citep{rathborne14b}. Because the highest volume density peaks have the shortest dynamical time-scales, they collapse and form stars first. As a result, the young stars initially inherit the hierarchical structure of the ISM, before attaining a spherically symmetric morphology through violent relaxation if the stellar group is gravitationally bound \citep[e.g.][]{gouliermis14}. This picture is supported by the fact that no starless clouds are known to exist in the Milky Way disc with properties similar to YMCs \citep[$M>10^4~\msun$ and $R<3~\pc$, see][]{ginsburg12,longmore14}, indicating that there are no candidate clouds for the in-situ formation of massive clusters. Even in the high-pressure environment of the Galactic Centre, which is more representative of the conditions of GC formation than the Galactic disc \citep{swinbank11,shetty12,kruijssen13c} and hosts the densest GMCs in the Milky Way \citep{longmore13b,rathborne14}, none of the GMCs have central densities as high as seen in YMCs \citep{walker14}.

The above results show that YMCs must acquire their mass through hierarchical growth, unless YMC formation can also proceed by other mechanisms that do not manifest themselves in the Milky Way. It has also been proposed that YMCs form through monolithic collapse to a very compact ($R\ll1~\pc$) protocluster and subsequently expand by gas expulsion \citep[e.g.][]{banerjee13}. However, this is inconsistent with the fact that all YMCs observed to date are seen to be in virial equilibrium, showing no signs of expansion following gas expulsion despite their young ages \citep[e.g.][]{rochau10,cottaar12,henaultbrunet12,clarkson12}. This cannot be caused by rapid revirialisation on a few-$\myr$ time-scale unless YMCs have initial densities much higher than ever observed \citep{banerjee13,longmore14}. Similar indications of hierarchical star and cluster formation are seen in external galaxies, where the structure of star-forming regions down to $10$--$20~\pc$ scales is hierarchical \citep{elmegreen01,bastian07,gouliermis10} and is subsequently erased on a galactic dynamical time \citep{gieles08c,bastian09b,bastian11}. These results show that {\it nearby YMCs assemble their mass through the hierarchical merging of smaller structures}.

\subsubsection{Young massive cluster formation time-scales} \label{sec:ymctime}
An important question is how long the YMC formation process proceeds. \citet{efremov98} showed that the age spread measured in a hierarchical distribution of star formation is largely a relic of the size-scale probed. However, star formation histories have also been derived for the centrally condensed, high-density clusters residing in the centres of more extended hierarchies. \citet{longmore14} provide an extensive literature review, finding that the age spreads observed in YMCs are all less than a few $\myr$. Using a new technique that is based on the age-dating of pre-main sequence stars, recent work has reported age spreads up to $\gtrsim10~\myr$ \citep[e.g.][]{beccari10,demarchi13}. However, such large age spreads are not seen in main-sequence or post-main sequence stars, suggesting that the large age spreads could be a byproduct of the new, pre-main sequence technique \citep{longmore14}.

In addition, all known YMCs in the Milky Way \citep[with ages $\tau>1~\myr$, cf.][]{portegieszwart10} are devoid of dense gas capable of forming stars -- as mentioned in \S\ref{sec:hierarchy}, those YMCs with observed kinematics are found to be in virial equilibrium, showing no signs of expansion following gas expulsion despite their young ages. Similar results are found in external galaxies, ranging from quiescent dwarfs to high-density galaxy mergers, where no evidence for ongoing star formation (traced by the H$\beta$ and O[{\sc iii}] lines) is seen for YMCs with masses $M=10^4$--$10^8~\msun$ and ages $t=10^7$--$10^9~\yr$ \citep{larsen11,bastian13b,cabreraziri14}. These results indicate that {\it the YMC formation process is likely finished in less than a $\mathit{Myr}$} (this estimate does not include the surrounding association or other, nearby sub-clusters), which at the volume densities of YMCs corresponds to one or two dynamical times \citep[cf.][]{elmegreen00}. Mass growth by `dry' mergers (i.e.~the violent relaxation of stellar sub-clusters) may continue for a few $\myr$ if the structure is gravitationally bound on a scale larger than the central stellar cluster \citep[e.g.][]{sabbi12}.

\subsubsection{The fraction of star formation occurring in bound stellar clusters} \label{sec:cfe}
Within the hierarchy of star-forming structures, the highest-pressure and highest-density peaks achieve the highest star formation efficiencies (SFEs) on the shortest time-scales, before star formation is halted by feedback \citep{elmegreen97,elmegreen08,kruijssen12d,wright14}. These high-density regions are the relevant environments for putting constraints on the formation of YMCs and GCs, because cluster formation models show that a high SFE is required to allow a stellar group to remain gravitationally bound upon gas removal \citep{tutukov78,hills80,lada84,boily03,moeckel10,pelupessy12,kruijssen12}. Consequently, only some fraction of all star formation results in bound stellar clusters. This fraction, often referred to as the cluster formation efficiency (CFE or $\Gamma$, see \citealt{bastian08}) has been found to be $\Gamma\sim0.05$ for star formation in the solar neighbourhood \citep{lada03}. However, recent work has suggested that this fraction is environmentally dependent, increasing from $\Gamma=0.01$ in quiescent dwarf and gas-poor spiral galaxies to $\Gamma>0.4$ in nearby starbursts \citep{larsen00,bastian08,goddard10,adamo10,adamo11,silvavilla11,cook12,ryon14}, indicating a trend of increasing CFE with the gas pressure (or surface density). Until recently, this trend could have been caused by the combination of inhomogeneous data sets and the fact that starburst galaxies are generally more distant, possibly leading to less well-resolved cluster populations and a spurious increase of the CFE. A recent paper by \citet{silvavilla13} may have settled the issue by presenting a variation of the CFE with galactocentric radius in the nearby, gas-rich spiral galaxy M83. Using a homogeneous data set, these authors find a systematic decrease of the CFE with the radius (and hence an increase with the gas surface density, which dominates over the changes of the other parameters), from $\Gamma=0.04$ at $R>2.7~\kpc$ to $\Gamma=0.15$ at $R\sim1~\kpc$. Adding a central value from \citet{goddard10}, the central CFE may be as high as $\Gamma\sim0.3$. These results show that {\it high gas pressures at the epoch of GC formation likely caused an elevated CFE compared to that seen in nearby galaxies}. In \S\ref{sec:phys}, it will be discussed how the environmental variation of the CFE should be expected from simple theoretical considerations.

\subsubsection{The initial cluster mass function} \label{sec:icmf}
Over the past two decades, observations of young stellar cluster populations have shown that {\it the initial cluster mass function (ICMF) satisfies a power law with index $\alpha=-2$ (${\rm d}N/{\rm d}M\propto M^{\alpha}$) across the full spectrum of host galaxies}, from quiescent dwarf galaxies to major mergers \citep[e.g.][]{zhang99,bik03,hunter03,mccrady07,larsen09,chandar10b,portegieszwart10}. Integration of this ICMF shows that the stellar mass is distributed evenly across young clusters of all masses, i.e.~the total mass per decade in mass is constant. While the universality of the ICMF slope is firmly established, some discussion remains on the mass range over which it holds. Because $\alpha=-2$, the ICMF must change slope or become truncated at both the low-mass and the high-mass end, because otherwise the total mass diverges. Next to this mathematical inconvenience, there are also physical reasons to expect a truncation. At the low-mass end, the fiducial truncation of $M_{\rm min}\sim100~\msun$ \citep[e.g.][]{lada03,lamers05b} may correspond to the mass-scale on which two-body relaxation-driven evaporation occurs on a star formation time-scale \citep{moeckel12}, implying that lower-mass clusters may never emerge visibly from the embedded phase. Unfortunately, the observational determination of the lower mass limit can only be carried out in the limited environment of the solar neighbourhood, beyond which $M\sim100~\msun$ clusters fall below the detection limit of modern telescopes. It is therefore unknown to what extent the fiducial lower mass limit is universal. The impact on GC formation studies is limited though -- assuming that the maximum cluster mass $M_{\rm max}>10^6~\msun$ (as is observed, see below), the extrapolation of observed ICMFs down to some value of $M_{\rm min}=10^1$--$10^3~\msun$ translates to an uncertainty on the total mass in clusters of less than $25\%$.

\subsubsection{The maximum cluster mass} \label{sec:mtoomre}
\begin{table}
 \centering
  \begin{minipage}{147mm}
   \caption{Adopted galaxy properties for the maximum cluster mass-scale comparison of Figure~\ref{fig:mtoomre}.}\label{tab:mtoomre}
  \begin{tabular}{@{}l c c c c c c c@{}}
\hline
System & $\Sigma$ & $\sigma$ & $\Gamma$ & $M_{\rm GMC,max}$ & $M_{\rm c}$ & $M_{\rm max}$ & References \\
   &  $[\msun~\pc^{-2}]$  &  $[\kms]$   &    &   $[\msun]$   &    $[\msun]$    &   $[\msun]$   &  \\[.2ex]\hline\\[-2.5ex]
Nearby spirals           &   $20$       &    $10$   &    $0.1$     &    $10^7$               &    $2\times10^5$   &    $6\times10^5$   &  1,2,3    \\
Antennae galaxies   &    $10^3$  &    $40$   &    $0.4$     &    $10^8$               &    $10^6$                &    $3\times10^6$   &   1,4,5,6   \\
M83 (inner field)       &    $40$      &    $15$   &     $0.1$    &    $3\times10^7$  &    $1.6\times10^5$ &    $2\times10^5$   &   7,8,9 \\
M83 (outer field)       &     $20$     &     $10$  &     $0.04$  &    $3\times10^7$ &    $5\times10^4$   &    $8\times10^4$   &   7,8,9 \\[.2ex]\hline\\[-2.4ex]
 \end{tabular}
 \footnotesize{References: (1) \cite{larsen09}, (2) \cite{bolatto08}, (3) \cite{kruijssen12d}, (4) \cite{wei12}, (5) \cite{gao01}, (6) \cite{schulz07}, (7) \cite{bastian12}, (8) \cite{lundgren04b}, (9) \cite{lundgren04}. The `nearby spirals' sample includes 22 spiral and five irregular galaxies and is listed in Table~1 of \citet{larsen09}. Note that the CFE for the Antennae galaxies has not been measured directly, but is based on similar, nearby starbursts, which have $\Gamma\sim0.4$ \citep{adamo11}. The inner and outer fields of M83 correspond to median galactocentric radii of $R=2.5~\kpc$ and $R=4.75~\kpc$, respectively \citep{bastian12}. For all galaxies, a universal GMC-scale SFE is assumed of $\epsilon=0.05$ \citep[e.g.][]{lada03}, which is necessarily lower than the SFE reached in the density peaks where the formation of bound stellar clusters takes place.}
\\
 \end{minipage}
\end{table}
At high masses, there should also be a physical truncation of the ICMF, which is often parametrized as an exponential truncation \citep{schechter76}, i.e.:
\begin{equation}
\label{eq:icmf}
\frac{{\rm d}N}{{\rm d}M}\propto M^{\alpha}\exp{(-M/M_{\rm c})} ,
\end{equation}
where $M_{\rm c}$ represents the characteristic truncation mass and $\alpha=-2$ as before. In stark contrast with the observational difficulty of measuring a minimum cluster mass, massive clusters can be readily observed out to large distances. However, the issue at the high-mass end is that the $\alpha=-2$ power law-shape of the ICMF yields ten times fewer clusters for every decade gained in mass. As a result, the maximum cluster mass should increase with the total mass of the cluster population or the star formation rate (SFR; given some characteristic cluster lifetime) solely due to small-number sampling statistics at high masses \citep{bastian08}. Probing the physical, high-mass truncation of the ICMF therefore requires rich cluster populations, in which size-of-sample effects are not important \citep{larsen02}. In recent years, several papers have presented evidence for such a truncation \citep{gieles06b,larsen09,bastian12,konstantopoulos13}, finding characteristic truncation masses across a broad range of $M_{\rm c}=0.5$--$10\times10^5~\msun$.

While it has not been shown explicitly, it is physically plausible that the ICMF truncation mass is related to the maximum GMC mass in galaxy discs, i.e.~the two-dimensional Jeans mass or Toomre mass \citep{toomre64}. This mass-scale reflects the maximum mass below which the self-gravity of GMCs can overcome the galactic differential rotation in an equilibrium disc \citep[e.g.][]{elmegreen83b,kim01}:
\begin{equation}
\label{eq:mtoomre}
M_{\rm T}=\frac{\sigma^4}{G^2\Sigma} ,
\end{equation}
where $\sigma$ is the gas velocity dispersion, which in galaxies across cosmic history ranges from a few~$\kms$ to several~$10^2~\kms$, and $\Sigma$ is the gas surface density, which ranges from a few~$\msun~\pc^{-2}$ to more than~$10^4~\msun~\pc^{-2}$.\footnote{While the functional form of equation~(\ref{eq:mtoomre}) suggests otherwise, the Toomre mass typically increases with the gas surface density, because in equilibrium galaxy discs $\sigma$ increases with $\Sigma$ more steeply than $\sigma\propto\Sigma^{1/4}$ (the underlying dependence is that both quantities increase with the pressure). This is easily seen empirically from a back-of-the-envelope argument: combining the condition for marginal Toomre stability $\pi G\Sigma=\sqrt{2}\Omega\sigma$ with the Schmidt-Kennicutt \citep{schmidt59,kennicutt98b} and Silk-Elmegreen \citep{silk97,elmegreen97b} star formation relations ($\Sigma_{\rm SFR}\propto\Sigma^{1.4}$ and $\Sigma_{\rm SFR}\propto\Sigma\Omega$, respectively, where $\Omega$ is the angular velocity) yields $\Omega\propto\Sigma^{0.4}$ and hence $\sigma\propto\Sigma^{0.6}$.} If the Toomre mass sets the maximum mass-scale for GMCs, then the maximum cluster mass-scale is obtained through multiplication with the GMC-scale star formation efficiency (SFE) and the CFE:
\begin{equation}
\label{eq:mmax}
M_{\rm T,cl}=\epsilon\Gamma M_{\rm T},
\end{equation}
where $\epsilon$ indicates the SFE. It is unclear if $M_{\rm T,cl}$ corresponds to the ICMF truncation mass $M_{\rm c}$ or the maximum cluster mass $M_{\rm max}$ (which for well-sampled cluster populations should be $M_{\rm max}>M_{\rm c}$), but this can be verified by comparing $M_{\rm T,cl}$ to the observed $M_{\rm c}$ and $M_{\rm max}$.

\begin{figure}
\center\includegraphics[width=12cm]{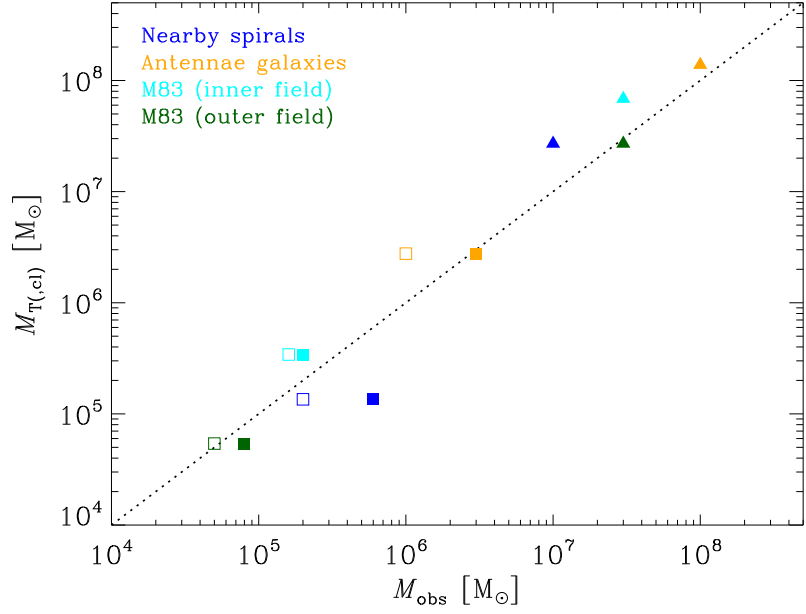}
\caption{This figure shows that the maximum masses of stellar clusters are related to the maximum mass-scale for gravitational instability in galaxy discs (the two-dimensional Jeans mass or Toomre mass). Shown are the predicted maximum GMC and YMC mass-scales ($M_{\rm T}$ and $M_{\rm T,cl}$, respectively) as a function of the observed maximum mass-scales $M_{\rm obs}$. Triangles indicate the Toomre mass $M_{\rm T}$ (see the text) as a function of the observed maximum GMC mass $M_{\rm GMC,max}$, whereas squares show the predicted maximum YMC mass $M_{\rm T,cl}=\epsilon\Gamma M_{\rm T}$ as a function of the observed maximum YMC mass $M_{\rm max}$ (filled squares) or the observed exponential truncation mass of the ICMF $M_{\rm c}$ (open squares). Observed YMC masses were obtained with stellar population modelling (see the references in Table~\ref{tab:mtoomre}). Colors indicate the host galaxies as indicated by the legend and the dotted line shows the 1:1 relation.} \label{fig:mtoomre}
\end{figure}
Although the overlap between observational samples of cluster populations and GMCs is still quite limited, it is large enough to make a first, qualitative comparison between (1) the observed maximum GMC mass $M_{\rm GMC,max}$, ICMF truncation mass $M_{\rm c}$ and maximum cluster mass $M_{\rm max}$, and (2) the estimated Toomre mass $M_{\rm T}$ and the expected maximum cluster mass $M_{\rm T,cl}$. Table~\ref{tab:mtoomre} lists the four examples that are considered here, and Figure~\ref{fig:mtoomre} shows a comparison of the different maximum mass-scales. While a simple Toomre argument may somewhat overestimate the maximum GMC and YMC mass-scales at high surface densities (i.e.~the Antennae and the inner field of M83), overall it does surprisingly well in predicting these quantities. The agreement with $M_{\rm max}$ is marginally better than with $M_{\rm c}$, but the difference between both mass-scales is only a factor of two. The encouraging implication is that {\it the masses of the most massive clusters may be used to infer the properties of their formation environment}, which provides a potential way of constraining the physics of GC formation (see e.g.~\citealt{harris94} and \S\ref{sec:pic}).

\subsubsection{Young massive cluster radii} \label{sec:rad}
YMCs have typical half-mass radii in the range $r_{\rm h}=0.5$--$10~\pc$, with little dependence on the cluster mass \citep{portegieszwart10}. \citet{larsen04b} introduced the relation $r_{\rm h}=3.75~\pc~(M/10^4~\msun)^{0.10\pm0.03}$ by fitting masses and radii to a large sample of extragalactic clusters with masses $10^3<M/\msun<10^6$ and ages $1<\tau/\myr<10^3$. This relation has been commonly used since, but the typical variation of the radius in a 1~dex mass interval is comparable to (or larger than) the systematic increase over the same interval, indicating that no strong mass-radius relation exists. \citet{portegieszwart10} find that the smallest radii reached by the youngest ($\tau<10~\myr$) clusters in the local Universe are set by a maximum half-mass density of $\rho_{\rm h}\sim5\times10^4~\msun~\pc^{-3}$ (and hence $r_{\rm h,min}\propto M^{1/3}$, also see \citealt{hopkins10}). However, the cluster sample is quite limited and it is uncertain how generally a maximum volume density would hold. The upper bound on the cluster radii in the $\tau<10~\myr$ sample of \citet{portegieszwart10} may either be consistent with a constant radius or a constant density, depending on the criterion used to define bound clusters. For ages $\tau>10~\myr$, {\it the sample is consistent with a constant radius of a few~$\pc$ irrespective of the definition used}.\footnote{\citet{pfalzner13} recently reported a second sequence of `clusters' in the mass-radius plane, with much larger radii than those considered here. The difference arises simply because we make a distinction between gravitationally bound and unbound systems. The large-radius sequence of \citet{pfalzner13} is constituted by unbound associations \citep[e.g.][]{blaauw64,wright14}, which evolve through ballistic expansion rather than the relaxation processes relevant for gravitationally bound clusters (and also for GC progenitors).}

Partly due to the substantial variation of cluster radii, the dependence of the initial YMC radius on the galactic environment is ill-understood. The present evidence for an environmental dependence is necessarily anecdotal due to the small sample size -- in the sample of \citet{portegieszwart10}, the Arches cluster is the densest (but not the most massive) YMC in the Milky Way, with $r_{\rm h}=0.5~\pc$ and $M=2\times10^4~\msun$, and resides in the strong tidal field of the Galactic Centre, which results in a tidal radius an order of magnitude smaller than it would have been in the solar neighbourhood. This difference is roughly consistent with the difference in radius between the Arches and YMCs in the Galactic disc, which have a median radius of several~$\pc$ in the \citet{portegieszwart10} sample. While a relation between the YMC radius and the local tidal radius may be plausible, more empirical tests are clearly needed to characterise the environmental variation of YMC radii.

\subsubsection{Conditions for globular cluster formation} \label{sec:conditions}
As will be discussed in \S\ref{sec:hiz}, the initial masses of all surviving GCs may have exceeded $10^5~\msun$. The discussion in this section shows that the formation of such massive clusters requires specific environmental conditions. For a galaxy to form a sizable population of possible GC progenitors, the maximum cluster mass should be $M_{\rm max}>{\rm several}~10^5~\msun$ and hence the Toomre mass should satisfy $M_{\rm T}>{\rm several}~10^7~\msun$. In the local Universe, such extreme conditions are only found in the high-pressure ($P/k\sim10^7~{\rm K}~{\rm cm}^{-3}$), vigorously star-forming environments of inner galactic discs, galaxy mergers and merger remnants, with turbulent velocity dispersions of $\sigma>20~\kms$ and gas densities of $\Sigma>10^2~\msun~\pc^{-2}$ (cf.~equation~\ref{eq:mtoomre}). Indeed, these systems host the most massive YMCs observed to date \citep[$M=10^7$--$10^8~\msun$, see e.g.][]{miller97,degrijs03,bastian06}. Massive cluster formation may be further promoted in high-pressure environments by the elevated cluster formation efficiencies observed under these conditions. Although a constant SFE was assumed in this section, it may increase with pressure too \citep{elmegreen97}. Finally, the associated high volume densities (up to two orders of magnitude higher than in local GMCs, see e.g.~\citealt{downes98}) imply that dynamical times of the order $0.5~\myr$ are already reached on the scales of GMCs, well before the formation of compact YMCs, suggesting that YMC formation in high-pressure environments ($P/k\sim10^7~{\rm K}~{\rm cm}^{-3}$) may proceed even faster than in nearby spiral galaxies like the Milky Way (where $P/k=10^4$--$10^5~{\rm K}~{\rm cm}^{-3}$).

\subsection{High-redshift galaxies} \label{sec:hiz}
The preceding section shows that some properties of stellar cluster populations are universal (such as the slope of the ICMF), whereas others depend sensitively on the galactic environment (such as the CFE and the ICMF truncation mass). Most GCs have ages indicating that they must have formed at $z>2$, i.e.~overlapping with the peak of the cosmic star formation history at $z=2$--$3$ \citep{hopkins06b}, when the environmental conditions in galaxies differed substantially from nearby galaxies in terms of their gas fractions, densities, kinematics, and pressures \citep{tacconi08,tacconi10,swinbank11,kruijssen13c}. This section therefore considers the gas and star formation properties of the high-redshift galaxies in which GCs may have formed, highlighting the implications for massive cluster formation.

\subsubsection{Galaxy properties and plausible massive cluster formation sites} \label{sec:sites}
Galaxies observed at $z>2$ have baryonic gas mass fractions $f_{\rm gas}>0.2$ (with a mean $f_{\rm gas}\sim0.5$ for stellar masses $M_{\rm star}>4\times10^9~\msun$, see \citealt{tacconi13}) and often show clumpy morphologies, indicating galaxy discs with high Toomre masses and strong gravitational instabilities \citep{elmegreen05,genzel11,tacconi13}. Observational surveys of gas-rich galaxies in the redshift range $z=0$--$3$ find a very limited variation of the molecular gas depletion time ($t_{\rm depl}\equiv M_{{\rm H}_2}/{\rm SFR}\sim1~\gyr$) in galaxies \citep{bigiel11,schruba11,tacconi13,fisher14}. As a result, the high gas fractions (i.e.~a high gas-to-stellar mass ratio $M_{{\rm H}_2}/M_{\rm star}$) necessarily imply a high specific SFR (${\rm sSFR}\equiv{\rm SFR}/M_{\rm star}=M_{{\rm H}_2}/M_{\rm star}t_{\rm depl}$), which indicates that {\it the gas-rich galaxies observed at $z>2$ are in the process of forming a large fraction of their (eventual) stellar mass}. The specific SFR is a function of galaxy mass and redshift -- following the ${\rm SFR}$--$M_{\rm star}$ relation in the redshift range $z=0$--$2$ of \citet{bouche10}, we can define a power-law parameterization for the specific SFR:
\begin{equation}
\label{eq:ssfr}
{\rm sSFR}\equiv\frac{{\rm SFR}}{M_{\rm star}}=\frac{M_{{\rm H}_2}}{M_{\rm star}}t_{\rm depl}=1.5~\gyr^{-1}\left(\frac{M_{\rm star}}{10^{11}~\msun}\right)^{-0.2}\left(\frac{1+z}{3.2}\right)^{2.7} ,
\end{equation}
which shows that low-mass, high-redshift galaxies have higher gas fractions and assemble a larger fraction of their stellar mass per unit time than high-mass, low-redshift galaxies. The star formation in these discs proceeds at high densities and pressures, with gas surface densities $\Sigma=10^2$--$10^{3.5}~\msun~\pc^{-2}$, SFR surface densities $\Sigma_{\rm SFR}=10^{-1}$--$10^{0.5}~\msun~\yr^{-1}~\kpc^{-2}$, and gas velocity dispersions $\sigma=10$--$100~\kms$ \citep{genzel10,swinbank12,tacconi13}, implying pressures of $P/k>10^7~{\rm K}~\cmc$.

The relation between galaxy mass and metallicity shows a clear redshift dependence, shifting down by a factor of 3--4 between the local Universe \citep[$z\sim0.1$,][]{tremonti04} and $z\sim2$ \citep{erb06}, and another factor of 2 between $z=2$ and $z=3$--$4$ \citep{mannucci09}. Observationally, the mass-metallicity relation is expressed as a variation of $\oh$ with galaxy mass, in contrast with the convention to express metallicities in terms of $\feh$ in GC studies. Following \citet{shapiro10}, we therefore assume $\oh\approx\zh=\feh+0.94\afe$ \citep{thomas03} with $\afe\sim0.3$. The mass-metallicity relations of \citet{erb06} and \citet{mannucci09} thus yield
\begin{equation}
\label{eq:fehmstar}
\feh \sim -0.59 + 0.24 \log{\left(\frac{M_{\rm star}}{10^{10}~\msun}\right)} - 8.03\times10^{-2} \left[\log{\left(\frac{M_{\rm star}}{10^{10}~\msun}\right)}\right]^2- 0.2(z-2) ,
\end{equation}
for the redshift range $z=2$--$4$ and assuming $12+\log{({\rm O}/{\rm H})}_\odot=8.7$ \citep{asplund09}. At a fiducial redshift of $z\sim3$, this enables the formation of $\feh\sim\{-1.6,-0.5\}$ clusters in $M_{\rm star}\sim\{10^8,10^{11}\}~\msun$ galaxies, respectively (cf.~\S\ref{sec:feh} below).

The clumpy discs of high-redshift galaxies may provide a natural environment for GC formation. It was proposed by \citet{shapiro10} that the observed dense and massive gas clumps ($R=1$--$3~\kpc$, $M=10^8$--$10^{9.5}~\msun$) may be the formation sites of several hundreds of (metal-rich) GCs, of which roughly a dozen survive to $z=0$. These clump mass-scales correspond to the extreme end of the spectrum of Toomre masses observed in the local Universe, indicating that high-redshift galaxies indeed satisfy the conditions required for the formation of massive clusters in general, and the formation of GC progenitors in particular. Note that the susceptibility of gas-rich discs to gravitational instabilities increases with the galactocentric radius \citep[e.g.][]{swinbank12}, and hence the formation of massive and dense, star-forming clumps should typically be restricted to radii larger than the half-light radius of the galaxy.

Considering the broad range of GC metallicities (see \S\ref{sec:feh} below), GC formation likely proceeded across the galaxy mass range. At the low-mass and low-metallicity end, plausible candidates are (damped) Ly$\alpha$ emitting systems \citep{burgarella01,elmegreen12}, whereas at high masses and metallicities, the clumpy and gas-rich discs of $z>2$ galaxies are a natural birth environment \citep{shapiro10}.

\subsubsection{A high cluster formation efficiency compared to nearby galaxies} \label{sec:cfez}
There are no empirical constraints on the CFE in $z>2$ galaxies. However, given the conditions under which star formation proceeds in these galaxies, it is possible to extrapolate the relation describing the CFE in local galaxies (see \S\ref{sec:cfe}) to the high-pressure regime. Although the main physical dependence of the CFE on the host galaxy properties is one on the pressure (and thus gas surface density, see \S\ref{sec:phys} and \citealt{kruijssen12d}), the popular empirical scaling relation connects the CFE observed in nearby galaxies with their SFR surface densities \citep[e.g.][]{goddard10,adamo11,silvavilla11}. Due to the inclusion of galaxy mergers and starbursting dwarf galaxies, these local-Universe SFR surface densities cover a range of $\Sigma_{\rm SFR}=10^{-3.5}$--$10^{0.5}~\msun~\yr^{-1}~\kpc^{-2}$, i.e.~almost reaching up to the upper end observed in high-redshift galaxy discs. At such extreme pressures, both the observed \citep{adamo11} and predicted \citep[Figure~9]{kruijssen12d} CFEs are roughly $\Gamma\sim0.5$, indicating that {\it in the high-pressure environments of $z>2$ galaxies, a five times larger fraction of all star formation results in bound stellar clusters than in nearby spiral galaxies}. Hence, the high-redshift conditions for star formation likely promote the formation of bound stellar clusters.

\subsubsection{A maximum mass-scale well in excess of globular cluster masses} \label{sec:mtoomrez}
Contrary to the CFE, the maximum mass-scale for gravitational collapse in galaxy discs (the Toomre mass, see~\S\ref{sec:mtoomre}) can be constrained empirically out to high redshift, owing to the high masses (and hence fluxes) of these clumps. \citet{swinbank12} show that the mass of the most massive clumps increases from several $10^7~\msun$ in nearby galaxies \citep{kennicutt03} to $M_{\rm T}\sim10^9~\msun$ at $z>2$ \citep[e.g.][]{genzel11}, in quantitative agreement with the simple Toomre argument made in \S\ref{sec:mtoomre}. Applying a typical CFE of $\Gamma=0.5$ and a SFE of $\epsilon=0.05$ yields a maximum YMC mass of $M_{\rm T,cl}\sim3\times10^7~\msun$, well in excess of the range of GC masses observed today. {\it The high-pressure environments at $z>2$ are able to accumulate sufficient mass to form GC-like YMCs}, contrary to most galaxies observed in the nearby Universe.

An important question remains: if the Toomre mass was as high as $M_{\rm T}>10^9~\msun$ in the most extreme galaxy discs at $z>2$, then why are there no GCs observed with masses $M>10^7~\msun$ (see \S\ref{sec:gcmf})? If the maximum cluster mass-scale is solely set by the Toomre mass, the $z>2$ environment may be somewhat {\it too} favourable for the formation of GC progenitors. Indeed, examples of YMCs with masses $M>10^7~\msun$ exist in local galaxies \citep[see e.g.~the compilation by][]{bastian13b}. Possible explanations for the lack of GCs with such high masses could be low total SFEs in high-redshift gas clumps due to stellar feedback, efficient dynamical friction of the most massive GCs towards the galactic nuclei, or rapid tidal disruption. Perhaps most importantly though, clump masses of $M_{\rm T}\sim10^9~\msun$ are the most extreme examples that have been observed to date. It is possible that integrated over the galaxy population, clumps at the lower end of the observable clump mass range ($M_{\rm T}\sim10^8~\msun$) are the most numerous, in which case the observed exponential truncation mass of the GC mass function ($M_{\rm c}\sim3\times10^6~\msun$, see \S\ref{sec:gcmf}) is entirely consistent (cf.~the Antennae galaxies in Table~\ref{tab:mtoomre} and Figure~\ref{fig:mtoomre}).

\subsubsection{Rapid early cluster disruption} \label{sec:hizdis}
A picture emerges in which the high-redshift environments of gas-rich, star-forming galaxies provide highly favourable conditions for the formation of massive stellar clusters. However, the formation of massive clusters does not guarantee their long-term survival. As will be highlighted below, the migration of YMCs out of their gas-rich, natal environments is essential for their survival until $z=0$.

In the local Universe, there is an empirical relation between the maximum ages reached by stellar clusters and the density of their galactic environment, both through (1) the large-scale gravitational potential (and hence the tidal field) and (2) the ambient gas density. The first of these two dependences can be characterised as a relation between the cluster disruption time $t_{\rm dis}$ and the density enclosed by the cluster orbit (the `tidal' density $\rho_{\rm tid}$), where $t_{\rm dis}\propto\rho_{\rm tid}^{-1/2}\propto\Omega^{-1}$ \citep{lamers05}. This scaling was determined for populations of centrally concentrated clusters in nearby galaxies spanning approximately an order of magnitude in tidal density, from the low-density environment of the Magellanic clouds ($\rho_{\rm tid}\sim10^{-2}~\msun~\pc^{-3}$), to the high tidal density of M51 ($\rho_{\rm tid}\sim0.2~\msun~\pc^{-3}$). This environmental dependence of the disruption time on the galactic environment is well-understood theoretically\footnote{There have been studies suggesting that cluster disruption time-scales are universal and do not depend on the galactic environment \citep[or the cluster mass, see e.g.][]{whitmore07,chandar10}. Because a wide range of theoretical literature has shown that cluster disruption is expected to be strongly environmentally dependent (see \S\ref{sec:phys}), this would be a important discovery. However, other observational work has shown the environmental independence disappears when accounting for differences in selection criteria \citep{bastian12,silvavilla14,fouesneau14}.} and is discussed in \S\ref{sec:phys} below.

The second environmental dependence of cluster disruption relates the disruption time to the strength and frequency of tidal shocking by encounters with molecular clouds. This disruption mechanism depends on the ambient gas density and strongly limits cluster lifetimes in gas-rich environments. In all galaxies with gas densities comparable to (or larger than) that of the Milky Way, tidal shocking poses a stronger limit to the cluster lifetime than the large-scale tidal field \citep{lamers06a,gieles06,elmegreen10b,kruijssen11}. Indeed, \citet{lamers05} find that the empirical scaling $t_{\rm dis}\propto\rho_{\rm tid}^{-1/2}$ only provides an upper limit to the cluster lifetime in gas-rich galaxies such as M51. Cluster disruption due to tidal shocks is discussed in more detail in \S\ref{sec:gasrich}.

The current evolution of GCs in galaxy haloes is deceptively quiescent when contrasting it with the likely, high-pressure formation environments of GCs at high redshift. While current mass loss rates of GCs are so low that some of their projected remaining lifetimes exceed $10^2~\gyr$ \citep{kruijssen09}, a $10^6~\msun$ cluster would survive for a mere $t_{\rm dis}\sim3~\gyr$ \citep{gieles05} in the gas-rich environment of M51 \citep[$\Sigma\sim30~\msun~\pc^{-2}$, e.g.][]{schuster07}. In gas-rich $z>2$ galaxies, this lifetime could be even shorter. In order for any GCs to have survived until $z=0$, the duration of the initial, rapid-disruption phase in their natal environment must have been limited, i.e.~{\it the young GCs must have escaped the gas-rich bodies of their host galaxies before they got destroyed}. An obvious physical agent for the galaxy-wide migration of stellar clusters would be the frequent galaxy mergers experienced by galaxies at high redshift. The galaxy merger rate may therefore be a key parameter in setting the properties of the surviving, present-day GC population. In \citet{kruijssen14c}, we quantify how the merger rate influences GC survival statistics such as the specific frequency (also see \S\ref{sec:tn}).

\subsection{Present-day GC properties} \label{sec:gc}
GCs have a wide variety of present-day properties that provide clues to the physics of their formation and subsequent evolution. Some of these are strongly correlated to the host galaxy properties (such as the specific frequency, i.e.~the number of GCs per unit stellar mass or luminosity, see e.g.~\citealt{peng08,forte14}), whereas others appear universal (such as the characteristic mass-scale, see e.g.~\citealt{harris91,strader06,rejkuba12}). In this subsection, we summarise the most important empirical constraints.

\subsubsection{The GC mass function} \label{sec:gcmf}
Contrary to the power-law mass function (${\rm d}N/{\rm d}\log{M}\propto M^{\alpha+1}$ with $\alpha=-2$) followed by young cluster populations forming in the Universe today \citep[see \S\ref{sec:ymc} and e.g.][]{zhang99,larsen09,chandar10b}, {\it the GC mass function (${\rm d}N/{\rm d}\log{M}$) is peaked, with a characteristic mass-scale of $M\sim10^5~\msun$} \citep{harris96,jordan07} {\it and a possible exponential truncation at the high-mass end}. In galaxies with masses similar to or higher than the Milky Way, the exponential truncation mass is $M_{\rm c}\sim3\times10^6~\msun$, see e.g.~\citealt{fall01,jordan07,kruijssen09b}.

Historically, the peaked mass function of GCs was explained by the special conditions of high-redshift star formation, such as the high Jeans masses expected after recombination \citep{peebles68} or in metal-poor gas in galactic haloes \citep{fall85}. However, HST's discovery of young clusters in nearby galaxies with masses exceeding those of GCs prompted a re-examination of these ideas -- the special, high-redshift conditions thought to be essential for GC formation proved to be unnecessary for the formation of massive clusters. The modern interpretation of the GC mass function is that it initially followed an $\alpha=-2$ power law, reflecting the hierarchical nature of the ISM from which the young GCs formed \citep{elmegreen97,fall10,kruijssen12b}, but was subsequently shaped by cluster disruption \citep{vesperini01,fall01}, which destroyed most of the low-mass clusters over the course of a Hubble time.\footnote{It has been suggested that the peaked shape of the GCMF is caused by the disruption of low-mass clusters by gas expulsion shortly after their formation \citep{parmentier07}. However, this model requires the mass function of the progenitor clouds to have a very high lower mass limit of $M_{\rm min}\sim10^6~\msun$ in high-pressure environments, which has not been observed \citep[e.g.][]{wei12}.}

\begin{figure}
\center\includegraphics[width=11cm]{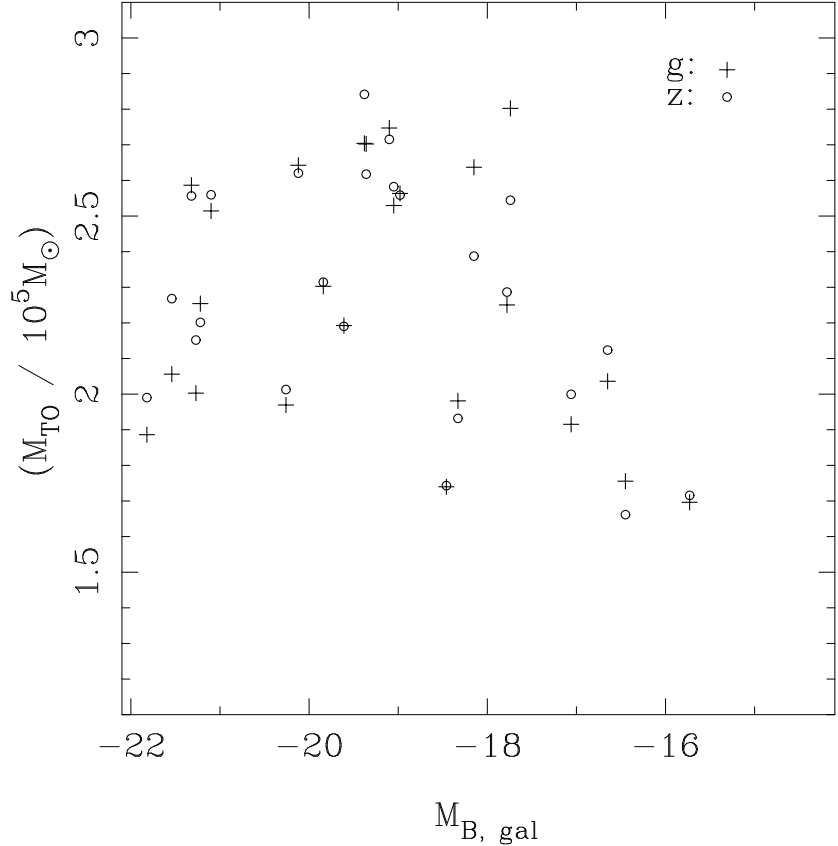}
\caption{This figure shows that the peak mass of the GCMF is near-universal, decreasing only slowly towards low-mass galaxies for $M_B>-20$. Shown is the GCMF peak mass ($M_{\rm peak}$ or $M_{\rm TO}$) as a function of the absolute $B$-band magnitude of the host galaxy $M_B$ for the galaxy sample of the ACS Virgo Cluster Survey. Plus symbols indicate the peak masses obtained from $g$-band GC luminosities, whereas open circles were obtained using $z$-band GC luminosities. Taken from \citet[Figure~15]{jordan07}, reproduced by permission of the AAS.} \label{fig:mpeak}
\end{figure}
However, a key problem remains. The characteristic mass of the GC mass function is remarkably universal -- as shown in Figure~\ref{fig:mpeak}, it is constant in Virgo Cluster galaxies with $B$-band magnitudes $M_B<-20$ and very weakly decreases with increasing magnitude in dwarfs with $M_B>-20$ \citep[at a rate of $\sim1$--$2\times10^4~\msun$ per magnitude,][]{jordan07}. Additionally, it is independent of the galactocentric radius within galaxies \citep[e.g.][]{vesperini03,mclaughlin08}. A large volume of literature has shown that cluster disruption is environmentally dependent \citep[see \S\ref{sec:hizdis} and][]{spitzer58,spitzer87,gnedin97,portegieszwart01,baumgardt03,kruijssen11,bastian12,silvavilla14,fouesneau14}, i.e.~the mass loss rate of GCs is set by the tidal field, which varies as a function of the galactocentric radius and the host galaxy. The relative universality of the GC mass function is therefore hard to reconcile with the variety of tidal fields GCs presently reside in \citep{vesperini03}. If cluster disruption is indeed responsible for the peaked GC mass function, the main question must be {\it when} GCs lost most of their mass -- and under which conditions. Because the formation of stellar clusters as massive as GCs requires specific environmental conditions (see \S\ref{sec:ymc} and \S\ref{sec:hiz}), it is not hard to imagine that the high-redshift formation sites of GCs were more similar than their wide variety of present-day environments, in which case early cluster disruption may have led to similar present-day mass functions \citep{elmegreen10,kruijssen12c}. Tentative evidence for the rapid evolution towards a common cluster mass-scale is found in nearby galaxy merger remnants \citep{goudfrooij04,goudfrooij07}. The mass loss history of GCs will be revisited in \S\ref{sec:phys}.

{\it The GC mass function has an (exponential) truncation at the high-mass end} \citep{harris14}{\it , with a strongly environmentally-dependent truncation mass} \citep{jordan07}, decreasing from $M_{\rm c}\sim3\times10^6~\msun$ in massive galaxies ($M_B<-20$) to $M_{\rm c}\sim3\times10^5~\msun$ in low-mass systems ($M_B>-18$). At even lower galaxy masses, the number of GCs becomes too small to draw statistically significant conclusions. This environmental variation of the exponential truncation mass is unexplained, but given the success of a simple Toomre mass argument in explaining the maximum masses of YMCs in nearby galaxies (see \S\ref{sec:mtoomre}), it is plausible that the same reasoning holds for the conditions of GC formation \citep[see e.g.][]{harris94,mclaughlin96}. The truncation could contribute to the near-universality of the GCMF peak mass, as it can greatly decelerate (or even stop) the increase of the peak mass due to disruption \citep{gieles09}.

\subsubsection{The GC metallicity distribution} \label{sec:feh}
\begin{figure}
\center\includegraphics[width=11cm]{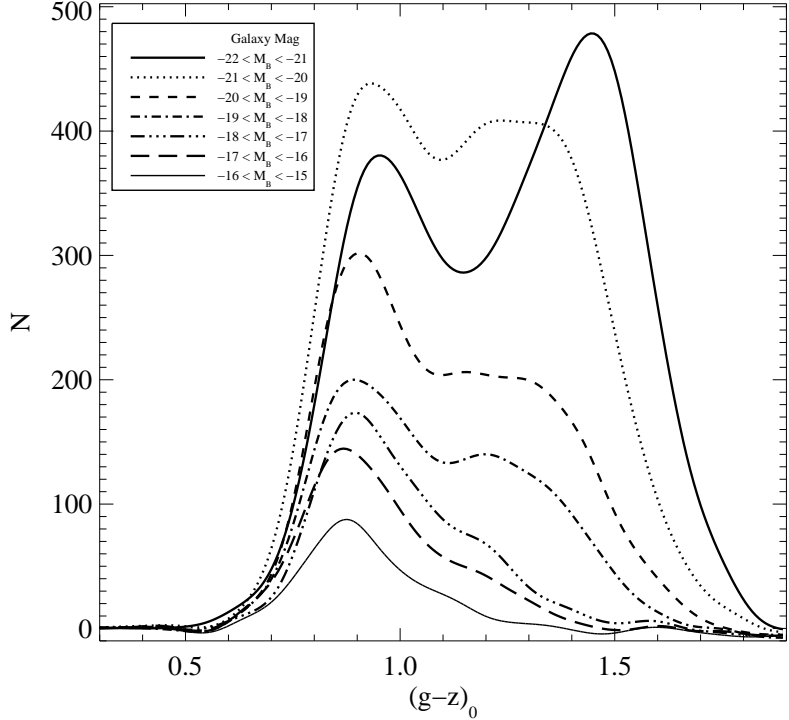}
\caption{This figure shows that the colour distribution of GCs is bimodal, with the relative strengths of the blue and red peaks depending on the host galaxy mass. Shown is the number of GCs per unit $g-z$ colour in seven bins of the absolute $B$-band magnitude of the host galaxy $M_B$ for the galaxy sample of the ACS Virgo Cluster Survey. As discussed in the text, the colour bimodality reflects and underlying metallicity bimodality, in which the red GCs are more metal-rich than the blue ones. Taken from \citet[Figure~5]{peng06}, reproduced by permission of the AAS.} \label{fig:bimodality}
\end{figure}

In massive galaxies like the Milky Way, {\it the metallicity ($\feh$) and optical colour (e.g.~$g-z$, $V-I$) distributions of GCs are bimodal}, with peak metallicities of $\feh\sim-1.6$ for metal-poor (blue) GCs and $\feh\sim-0.5$ for metal-rich (red) GCs \citep{kinman59,zinn85,forbes97,larsen01,bica06,peng06,brodie06,chiessantos12}. An example is given in Figure~\ref{fig:bimodality}, which shows the colour distributions of GCs in Virgo Cluster galaxies \citep{peng06}. The peak colours (and metallicities) depend weakly on the host galaxy mass, increasing from $\{(g-z)_{\rm blue},(g-z)_{\rm red}\}=\{0.85,1.1\}$ (corresponding to $\{\feh_{\rm blue},\feh_{\rm red}\}=\{-1.8,-0.8\}$) at $M_B>-16$ to $\{(g-z)_{\rm blue},(g-z)_{\rm red}\}=\{0.95,1.4\}$ (corresponding to $\{\feh_{\rm blue},\feh_{\rm red}\}=\{-1.4,-0.2\}$) at $M_B<-21$. The relative distribution of GCs over the metal-poor and metal-rich populations is a function of the galaxy mass too. In the example of Figure~\ref{fig:bimodality}, the metal-rich GC population is almost absent in galaxies with $M_B>-17$, whereas it starts to dominate over the metal-poor GC population in galaxies with $M_B<-21$. As a result, the mean GC metallicity increases more strongly with galaxy mass than the peak metallicities of the subpopulations (\citealt{peng06}, but also see e.g.~\citealt{brodie06}).

It has been shown in the literature that the translation from colour to metallicity is non-linear due to the presence of hot horizontal-branch stars \citep[e.g.][but also see \citealt{peng06}]{yoon06}, which could imply that the bimodality of the colour distribution results from the non-linear mapping of a unimodal metallicity distribution function to colour space. However, spectroscopic observations that directly probe the chemical composition of GCs are capable of constraining the shape of the GC metallicity distribution function independently from the conversion between colour and metallicity. Such observations of GCs in both the early-type galaxy NGC~3115 \citep{brodie12} and the Milky Way \citep{harris96} unequivocally show a metallicity bimodality.

Metal-poor GCs with masses $M>{\rm several}~10^5~\msun$ exhibit a relation between GC colour and magnitude (the `blue tilt'), with brighter GCs being redder \citep[e.g.][]{harris06}. It was shown by \citet{mieske10} that a similar, but much weaker relation may exist for metal-rich GCs. Both relations have been interpreted in the literature as being the combined result of the pre-enrichment of the progenitor GMC, through which more massive GCs may have become more enriched (i.e.~redder), GC self-enrichment, and dynamical evolution, by which lower-mass GCs have become bluer due to the loss of low-mass stars through evaporation \citep{bailin09,goudfrooij14}. However, these constraints on GC formation and enrichment are necessarily indirect and should be tested against other models, e.g.~using the star formation time-scale in GMCs to quantify the time available for pre-enrichment (also see \S\ref{sec:chem}).

Observations of GCs in the Milky Way and in other galaxies have shown that the metal-rich GC population is associated with the main body (bulge and/or disc) of the host galaxy in terms of its kinematics \citep[e.g.][]{zinn85,dinescu99} and metallicity \citep[e.g.][]{forbes97}. The spatial distributions of the GC population and coeval field stars are similar, but when including the younger field star populations the GC population is spatially more extended. By contrast, metal-poor GC populations are characterised by dispersion-dominated kinematics that are uncorrelated with those of the host galaxy and have metallicities lower than those of metal-rich GCs (and their host galaxies) by roughly an order of magnitude \citep{forbes97}. The spatial distributions of metal-poor GC populations extend well into the haloes of their host galaxies, further than those of metal-rich GC populations \citep[e.g.][]{harris09,harris09b,strader11,usher13,pota13}.

The dissimilar metallicities, kinematics and spatial distributions of the metal-poor and metal-rich GC populations have led to a surge of literature in which the bimodality between both populations is explained by different formation physics. For instance, metal-rich GCs were proposed to have formed in galaxy mergers \citep{ashman92,beasley02} or coevally with the bulk of the host galaxy's stellar population \citep{forbes97,strader05}, whereas metal-poor GCs were suggested to have formed during the collapse of protogalactic clouds \citep{forbes97,beasley02}, which was possibly truncated by reionization \citep{strader05}. 

Another explanation for the bimodality of GC populations is that it is a natural result of hierarchical galaxy formation across the galaxy mass and metallicity range, in which metal-poor GCs originated in cannibalised dwarf galaxies \citep[e.g.][]{cote98,hilker99,kisslerpatig00,mackey04}. This idea has gained a lot of traction in recent years. By embedding the redistribution of GCs during galaxy formation, new models quantitatively reproduce the metallicity bimodality seen in observed GC populations \citep{muratov10,tonini13,li14}, but they arrive at similar conclusions in different ways. Crucially, these models still rely on a number of ad-hoc assumptions regarding the initial GC population -- they all lack a self-consistent model for GC formation and need to assume that some fraction of GCs formed per unit host galaxy stellar mass (i.e.~the specific frequency). This fraction could be chosen to reflect the specific frequencies observed at $z=0$ without accounting for any evolution between GC formation and the present day \citep{tonini13}. Alternatively, the fraction of the galaxy mass constituted by GCs could be widely different at high redshift and may have attained its present-day value due to GC evolution \citep{muratov10,li14}. However, because models including GC evolution do not include the physics of GC formation, the description of GC evolution is necessarily restricted to the time during which they reside in the host galaxy halo, whereas the initial phase in high-pressure discs may be more important -- see \S\ref{sec:hizdis} and \S\ref{sec:phys}. These assumptions either need to be overcome or should be physically motivated before a theoretical understanding of GC formation is achieved.

Likewise, the metallicity-dependence of the spatial distribution of GCs has been used to constrain the formation redshifts of GCs \citep[e.g.][]{moore06,brodie06,boley09,moran14}. This is done using cosmological simulations -- at $z=0$, the spatial profiles of dark matter haloes that collapsed at redshift $z_{\rm coll}$ are matched to the spatial profiles of observed GC (sub-)populations. The best-fitting value of $z_{\rm coll}$ is then taken to be the formation redshift of GCs. This approach effectively attempts to match the binding energies of GC populations to those of cooling, high-redshift haloes, and omits the physics of GC formation and disruption. However, it is unclear if this assumption is valid -- can halo collapse be directly related to GC formation without accounting for baryonic physics? As is discussed at length throughout this review, the physics of GC formation and disruption determine which objects may survive to $z=0$ and what their spatial distribution will be. As a result, including physical models for GC formation and disruption is crucial to avoid ad-hoc results. To obtain results that are applicable to the observed GC population, combining GC models with cosmological simulations is a key target for the coming years.

\subsubsection{The specific frequency of GCs} \label{sec:tn}
The specific frequency $S_N$ of a galaxy's GC population was originally defined as the number of GCs per unit host galaxy luminosity, i.e.~$S_N=N_{\rm GC}\times10^{0.4(M_V+15)}$ with $N_{\rm GC}$ the number of GCs and $M_V$ the galaxy's absolute $V$-band magnitude \citep{harris91,kisslerpatig97,miller07}, introducing a measure of the `richness' of a GC population across the galaxy mass range. In a Universe characterised by hierarchical galaxy growth \citep{white78,white91}, the specific frequency is a useful quantity to trace the accretion and merging of GC populations, because it does not intrinsically depend on the host galaxy luminosity -- if a fixed fraction of a galaxy's mass resides in GCs, then then the number of GCs and the galaxy luminosity are linearly proportional to each other, resulting in a constant specific frequency. As a result, the specific frequency is a useful quantity to compare between different galaxy masses and (potentially) redshifts. However, the normalisation per unit galaxy luminosity is only meaningful if the mass-to-light ratio is constant across the galaxy population, i.e.~the variation of galaxy age and metallicity is limited. For this reason, \citet{zepf93} defined the specific frequency per unit stellar mass as $T_N=N_{\rm GC}(M_{\rm star}/10^9~\msun)^{-1}$, which is the definition of the term `specific frequency' that is used in this review. If GCs would have infinite lifetimes, the specific frequency would simply indicate which fraction of star formation resulted in GCs at the time of GC formation. However, GCs are disrupted over time, and therefore the specific frequency also factors in which fraction of GCs have survived until the present day.

Much like it has played a major role in characterising the GC mass function (\S\ref{sec:gcmf}) and the bimodality of GC colours and metallicities (\S\ref{sec:feh}), the ACS Virgo Cluster survey (ACSVCS) enabled the study of systematic variations of the specific frequency with host galaxy properties. \citet{peng08} showed that in Virgo Cluster galaxies $T_N$ monotonically decreases with increasing galaxy mass for a limited (but substantial) part of the galaxy mass range, from $T_N\sim30$ at $M_{\rm star}=3\times10^8~\msun$ to $T_N\sim5$ at $M_{\rm star}=3\times10^{10}~\msun$, before increasing again for galaxy masses $M_{\rm star}>3\times10^{10}~\msun$. In addition, \citet[also see \S5.1 of \citealt{peng08} and \citealt{lotz04}]{georgiev10} considered a sample of nearby dwarf galaxies and found that the trend of increasing specific frequency with decreasing galaxy mass at low galaxy masses may extend to galaxies less massive than the minimum galaxy mass in the ACSVCS of $M_{\rm star}\sim10^8~\msun$. However, the GC populations of galaxies with masses $M<10^8~\msun$ start suffering from substantial Poisson noise (i.e.~systems without any GCs at all become increasingly numerous), which may lead to a selection bias: when only considering galaxies that do host GCs (and hence $\lim_{M_{\rm star}\downarrow0}N_{\rm GC}=1$), one automatically finds $T_N\propto M_{\rm star}^{-1}$ for low galaxy masses.

Strong deviations from the above specific frequencies are found when comparing the GC population to a restricted sample of field stars with similar metallicities (assumed to trace the coeval stellar population). Recent papers by \citet{larsen12} and \citet{larsen14} have shown that in the nearby dwarf galaxies Fornax, IKN and WLM, the GCs constitute $\sim20\%$ of the metal-poor ($\feh<-2$) stellar mass. This number is uncertain by a factor of 2--3 due to assumptions regarding mass-to-light ratios and/or star formation histories, but regardless it provides a strong contrast with the typical $T_N\sim10$ seen when including the entire field star population in massive galaxies. Assuming a typical GC mass of $M\sim10^5~\msun$, the Larsen et al.~results suggest $T_N\sim2\times10^3$ for $\feh<-2$ in dwarf galaxies. Similar metallicity-resolved analyses of more massive galaxies are needed to infer if such extreme specific frequencies persist during hierarchical galaxy growth.

As outlined above, the specific frequency is an ideal quantity to characterise the GC population during hierarchical galaxy growth and may therefore provide clues to the formation environments of GCs \citep[e.g.][]{peng08,forte14}. For this reason, recent work has begun to correlate the total mass of galactic GC populations to the dark matter halo masses of their host galaxies \citep{peng08,spitler09,harris13,hudson14,durrell14}. These papers show that the total mass in GCs $M_{\rm GC,tot}$ is a roughly constant fraction $\eta$ of the halo mass $M_{\rm h}$, with $\log{\eta}\equiv \log{(M_{\rm GC,tot}/M_{\rm h})}=-4.2$ in the halo mass range $M_{\rm h}=10^9$--$10^{15}~\msun$ \citep{spitler09,harris13}.

When measuring the GC-to-halo mass ratio, it is important to acknowledge its radial variation and cover the entire halo out to the virial radius \citep{alamomartinez13}. This was recently done as part of the Next Generation Virgo Cluster Survey, which covers M87 and M49 out to the virial radius, resulting in a slightly lower, but surprisingly similar value of $\log{\eta}\equiv \log{(M_{\rm GC,tot}/M_{\rm h})}=-4.5$ (\citealt{durrell14}, also see \citealt{hudson14}). If this relation between dark matter halo mass and the total GC mass is fundamental, it could give rise to secondary correlations \citep{blakeslee97}, such as a constant, spatially resolved ratio between the total mass in GCs and the total baryonic mass in high-mass galaxies \citep{mclaughlin99}, or the relation between the number of GCs and the host galaxy's supermassive black hole mass \citep{burkert10}.

If the linear relation between total GC mass and dark matter halo mass indeed reflects a fundamental connection between GC formation and the collapse of dark matter haloes at high redshift, then the variation of specific frequency with galaxy mass reflects a variation of the field star formation efficiency with galaxy mass, rather than providing any insight into GC formation \citep[e.g.][]{peng08,spitler09}. This line of reasoning follows the early arguments made by \citet{peebles68} and \citet{fall85} that GC formation requires special conditions unique to the high-redshift Universe such as a high Jeans masses after recombination or in metal-poor gas (see \S\ref{sec:gcmf}). However, a fundamental relation between GCs and dark matter haloes at high redshift must have been modified by a Hubble time of cluster disruption, unless it transpired independently of the galactic environment, which is at odds with theory and observations \citep[see \S\ref{sec:phys} below and e.g.][]{fall01,lamers05,gieles08,elmegreen10,kruijssen11,bastian12}.

In order to assess how important the environmental variation of cluster disruption is to the relation between the specific frequency (or total GC mass) and host galaxy properties, the dependence of the specific frequency on the local galactic environment should be mapped. GC disruption in galaxy haloes depends primarily on the angular velocity ($t_{\rm dis}\propto\Omega^{-1}$) or, for a constant circular velocity, on the galactocentric radius ($t_{\rm dis}\propto R$). If GC disruption in their present-day halo environments would affect the specific frequency, one would therefore expect a relation between the specific frequency and the galactocentric radius within a single galaxy. In practice, this is less straightforward. The GC populations of massive galaxies exhibit a radial metallicity gradient, because the metal-poor population is spatially more extended than the metal-rich population (see \S\ref{sec:feh}). Since (1) galaxies form hierarchically from the merging of lower-mass galaxies, (2) metallicity increases with galaxy mass (see \S\ref{sec:hiz}), and (3) the specific frequency increases towards low galaxy masses, the radial metallicity gradient of GC populations implies that the specific frequency must increase with galactocentric radius. This trend would also be expected due to GC disruption \citep{lamers14}. Regardless of whether it is caused by disruption or by metallicity, observations show that the specific frequency indeed increases with galactocentric radius -- the radial profiles of GC populations are shallower than those of the host galaxy light \citep[e.g.][]{bassino06,harris09,strader11,kartha14}.

\begin{figure}
\center\includegraphics[width=12cm]{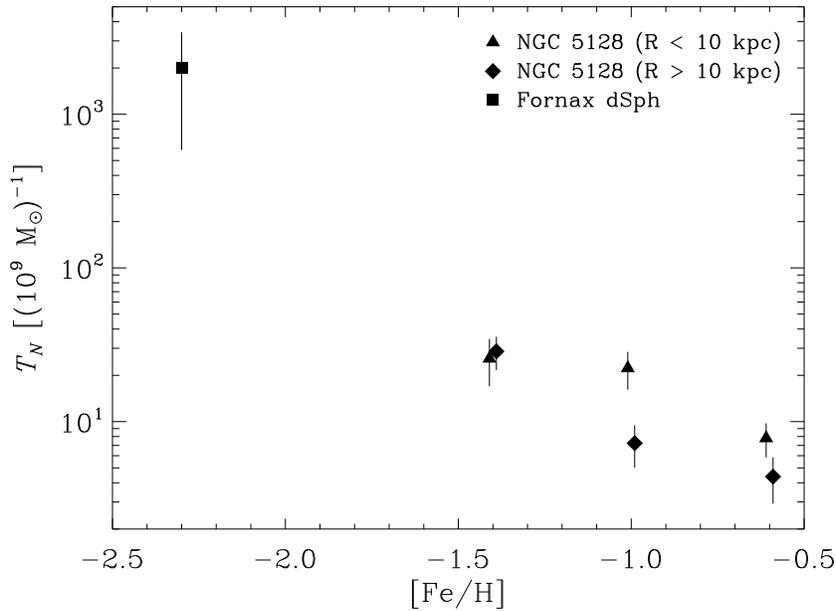}
\caption{This figure shows that the specific frequency of GCs does not depend on the galactocentric radius and must have been set during their formation or early evolution. Shown is the specific frequency $T_N$ as a function of metallicity $\feh$ for GCs in the inner region of NGC~5128 \citep[triangles, $R\lesssim10~\kpc$; data from][]{harris02}, the outer region of NGC~5128 \citep[diamonds, $R\gtrsim10~\kpc$; data from][]{harris02}, and in the Fornax dSph \citep[square; data from][]{larsen12}. The data points for GCs in NGC~5128 show that $T_N$ primarily depends on $\feh$ and any systematic variation of $T_N$ with the galactocentric radius is induced by the galaxy's radial metallicity gradient. The data point for the Fornax GC system is included to illustrate the metallicity trend to lower metallicities, but being a single point it does not provide further constraints on the radius-dependence of the specific frequency. This figure will appear in \citet{kruijssen14c}.} \label{fig:tnfeh_obs}
\end{figure}
To circumvent the radius-metallicity degeneracy, it is desirable to consider either the radial variation of the specific frequency at a given metallicity  \citep{forte07}, or the metallicity-dependence of the specific frequency in certain radial intervals \citep{harris02,lamers14}. The latter is shown in Figure~\ref{fig:tnfeh_obs} for GCs in the nearby early-type galaxy NGC~5128 \citep[Centaurus~A, using data from][]{harris02} and the Fornax dwarf spheroidal \citep{larsen12}. The data points from \citet{harris02} for $\feh<-1.6$ and $\feh>-0.4$ are omitted, because it is highly implausible that GCs and field stars would cover the same metallicity ranges in open-ended intervals. Figure~\ref{fig:tnfeh_obs} shows that in NGC~5128, the specific frequency decreases monotonically with metallicity, independently of the galactocentric radius (this will be discussed in more detail in \citealt{lamers14}). If anything, the specific frequency of metal-rich GCs is slightly higher at small radii than at large ones, contrary to what would be expected for cluster disruption. This means that the disruption of GCs in their present-day environments does not affect their specific frequency and hence must be limited. It also implies that models utilising the disruption of GCs in galaxy haloes for explaining present-day properties of GC populations should be critically evaluated, as they will not be able to reproduce the trend of Figure~\ref{fig:tnfeh_obs}.

Given the limited influence of recent disruption, it appears that the specific frequencies of GCs were set before they reached their present-day spatial distributions in galaxy haloes. Does this indicate that there is a fundamental relation between GCs and the formation of dark matter haloes after all? Or was the formation efficiency of GCs at high redshift environmentally dependent? It remains possible (or even plausible, see \S\ref{sec:pic}) that cluster disruption during the early evolution of young GCs in gas-rich environments affected the statistics of the eventual GC population. Irrespective of its reason, understanding the trend of Figure~\ref{fig:tnfeh_obs} is of critical importance in explaining the origin of GCs.

\subsubsection{Chemical abundance patterns} \label{sec:chem}
Until recently, GCs were thought to host single stellar populations, i.e.~all stars were thought to have the same age and initial chemical composition. While exceptions had already been known since the 1970s \citep{osborn71,smith87,kraft94}, this picture has changed particularly in the last few decades due to systematic observations of multiple stellar populations in GCs in terms of their chemical abundance patterns, extending the earlier observations of giant stars in GCs by also covering the main sequence \citep[see e.g.][]{gratton12}. The availability of these new observations has triggered the rapid growth of a highly specialised field aiming to constrain the physics of GC formation using observed chemical abundance variations within single GCs, mostly focussing on proton-capture elements such as C, N, O, F, Na, Mg, Al, and Si. There now exists a wide range of GC formation models aimed specifically at explaining the observed abundance patterns. \citet{gratton12} recently reviewed the observational side of the problem -- providing another update here would be unrealistic. Instead, we will briefly focus on some of the specific predictions and assumptions made by the main three models for the observed abundance variations, aiming to reconnect these to the constraints on GC formation gained from studies of nearby YMC formation, high-redshift galaxies, and present-day GC populations (\S\ref{sec:ymc}--\ref{sec:tn} above).

Observationally, the presence of multiple stellar populations in GCs manifests itself in a broad range of chemical abundance patterns and particular morphologies in the colour-magnitude diagram. Excellent summaries are presented by \citet{renzini08} and \citet[\S1]{bastian13}, from which the key points are given here. 
\begin{itemize}
\item[(i)]
Stars within a single GC exhibit an anti-correlation between their Na and O abundances, which implies a spread in light elements and indicates incomplete mixing of the enriched material. This has been proposed as a defining feature of GCs, as it has been found in all considered GCs \citep[$M\geq3\times10^4~\msun$, e.g.][]{mucciarelli09,carretta10}, but it is still unclear how generally it holds across the GC mass range.
\item[(ii)]
The colour-magnitude diagrams of a substantial subset of GCs show spreads in the main sequence and/or its turn-off \citep[e.g.][]{piotto07}, either due to He enrichment (which is a common, but not a universal property of enriched populations) or a C--N anti-correlation (and a spread of the C and N abundances).
\item[(iii)]
The enriched population shows no Fe enhancement, indicating no significant contribution from supernovae. In some rare cases, GCs show a spread or bimodality in their Fe abundance (e.g. $\omega$Cen, M22). These GCs have been hypothesised to be the remnants of dwarf galaxy nuclei or nuclear clusters \citep{zinnecker88,freeman93,lee09,pfeffer14}. Considering the ubiquitous GC mass-metallicity relation seen at high masses in extragalactic observations \citep[see \S\ref{sec:feh} and e.g.][]{mieske10}, it is possible that other formation mechanisms for Fe-enriched GCs exist too.
\item[(iv)]
It is generally found that the enriched stars are more centrally concentrated than the non-enriched stars. However, see \citet[NGC~6362]{dalessandro14} and Alonso-Garcia et al., in prep. (NGC~288) for two examples of the opposite. Additionally, \citet{larsen14b} have shown that the spatial distributions of the different sub-populations in NGC~7078 show a more complex distribution. It is currently unknown how representative this GC is.
\item[(v)]
GCs contain roughly equal numbers of chemically enriched and unenriched stars, but there exists substantial cluster-to-cluster variation \citep[e.g.][]{dantona08}.
\end{itemize}

Three scenarios have been proposed to explain the above (and other) abundance patterns, each of them with their own assumptions, requirements, and predictions. Each of these are summarised below. The first and second scenarios both require a second {\it generation} of stars to form early in a GC's life. The exact time at which this second generation may form ($\Delta\tau<200~\myr$) cannot be directly constrained observationally, because it is much smaller than the typical ages of GCs ($\tau\sim10^{10}~\yr$), with $\Delta\tau/\tau<0.02$. By contrast, the third scenario proposes that part of the stellar population is polluted through an accretion mechanism onto {\it preexisting} stars, avoiding the need for further star formation after a GC is born. Rather than describing the successes of these three models, the list below focusses on their problems and shortcomings, as these highlight the areas in which something new can be learned.
\begin{itemize}
\item[(i)]
{\it Enrichment by asymptotic giant branch (AGB) star winds} \citep[e.g.][]{ventura01,dercole08,conroy11}. In this model, a massive cluster forms from pristine gas and clears out the remaining gas. After the last type-II supernova has exploded (i.e.~after $\sim30~\myr$), the winds from AGB stars are too slow to escape the cluster's gravitational potential and they accumulate at the centre. For a \citet{chabrier03} initial mass function (IMF), 8\% of the stellar mass is initially locked up in AGB stars. In order to produce an enriched population of approximately equal size to the unenriched population, it is necessary to lose a large part of the unenriched population by cluster disruption (the {\it mass budget problem}). In addition, while different models relying on AGB ejecta exist, none of these reproduce the observed Na-O anti-correlation, but instead predict a correlation between both abundances. This can be solved by introducing (super-)AGB stars to produce high Na and low O abundances in the enriched material and subsequently diluting this by the accretion of low-Na, high-O, pristine material before the formation of the enriched population, which causes the Na abundance to decrease as the O abundance increases. The amount of accreted material must be as high as about half of the total GC mass despite possibly highly energetic stellar feedback (the {\it gas accretion problem}). This dilution scenario requires GCs to have been 10--20 times more massive at birth than they are today \citep[referring back to the mass budget problem, see e.g.][]{dercole08,conroy12}.
\item[(ii)]
{\it Enrichment by fast-rotating massive star (FRMS) winds} \citep[e.g.][]{decressin07,krause13b}. In this model, a mass-segregated, massive cluster forms from pristine gas and does not manage to clear out the remaining gas for $\sim30~\myr$ due to the deep gravitational potential, except in the direct vicinity of the massive stars. This allows the remaining gas to accrete onto the decretion discs surrounding the FRMSs, where it is enriched by the FRMS ejecta and cools to form an enriched stellar population. This phase is completed after $\sim10~\myr$, after which supernovae fail to eject any unused gas, but stir it up sufficiently to inhibit further accretion. After $\sim30~\myr$, the gas is finally expelled by heating from accreting stellar remnants. Note that this model does not require the accretion of pristine material, but instead relies on material left-over from the formation of the unenriched population and hence must assume a low local SFE, possibly at odds local YMCs, which (1) are already gas-poor a few~$\myr$ after their formation and (2) are gravitationally bound, suggesting a high local SFE (the {\it gas retention problem}, see \S\ref{sec:ymctime}). For a \citet{chabrier03} IMF, 3\% of the stellar mass is initially locked up in FRMSs. In order to produce an enriched population of approximately equal size to the unenriched population, it is necessary to lose a large part of the unenriched population by cluster disruption, requiring GCs to have been initially $>20$ times more massive than they are today (the {\it mass budget problem}).
\item[(iii)]
{\it Enrichment by the early disc accretion (EDA) of FRMS and massive interacting binary (MIB) winds} \citep{bastian13}. In this model, low-mass, pre-main sequence stars sweep up enriched material from the slow winds of FRMS and massive interacting binaries (MIBs, \citealt{demink09}) with their protoplanetary discs. Because these stars are still fully convective, material accreted from their discs is efficiently mixed.\footnote{\citet{dantona14} suggested that the accreting star may not remain fully convective for the entire duration of the accretion phase. However, this is not found in other models \citep[e.g.~BASTI,][]{pietrinferni06}, which predict a longer fully convective phase.} For a \citet{chabrier03} IMF, 13\% of the stellar mass is initially locked up in FRMSs and MIBs, but this material is only used to enrich low-mass stars, because they retain their protoplanetary discs much longer than high-mass stars. The model thereby solves the mass budget problem that affects the other two models, nor does it suffer from the gas accretion problem since it requires no pristine gas accretion. However, it does require protoplanetary discs around low-mass stars to survive for 5--10~$\myr$, which could be (marginally) at odds with disc lifetimes estimated for local YMCs \citep{haisch01,bell13}, in which discs are disrupted by dynamical encounters \citep[e.g.][]{rosotti14} and (internal and external) photoevaporation \citep[e.g.][]{scally01,adams04}, especially at the high stellar densities \citep[e.g.][]{johnstone98,dejuanovelar12} that are characteristic of GCs (the {\it disc lifetime problem}). A final issue of this model is that it predicts a strong depletion of Li in the enriched population, whereas it is observed to be constant or only mildly depleted. This problem arises for all models relying on enrichment by massive stars (hence it also affects the FMRS model, see \citealt{salaris14}) -- even the AGB model requires some fine-tuning of  the Li production rate in AGB stars by the Cameron-Fowler mechanism (which is a strong function of the mass loss rate) to balance the Li destruction rate \citep[e.g.][]{ventura10,gratton12}.
\end{itemize}

When attempting to reverse-engineer a possibly complex physical process from the relics left after a Hubble time of evolution, it is essential to critically evaluate the resulting theories in order to prevent their uninhibited growth and to minimise their complexity. Few aspects of GC formation physics can be tested using direct observations, but it is possible to compare the above models to nearby YMC formation -- none of the models invoke special, early-Universe physics, hence their predictions should also hold for (and be testable using) observations of YMC formation in the local Universe, provided that the conditions (e.g.~pressures and densities) are similar to those of GC formation -- as is the case in e.g.~galaxy mergers or starburst galaxies (see \S\ref{sec:ymc} and \S\ref{sec:hiz}).

As discussed at length in \S\ref{sec:ymc}, nearby YMCs show no evidence of star formation beyond the first few~$\myr$, which seems to rule out the AGB and FMRS models due to the gas accretion and gas retention problems. While YMCs do not display the chemical abundance variations seen in GCs either \citep{davies09,mucciarelli11,mucciarelli14}, the AGB and FMRS models do predict the same enrichment processes should be taking place in YMCs -- as stated, these models do not rely on special, early-Universe physics. The EDA model does predict a (strictly observational) difference between YMCs and GCs, because in this model only low-mass ($m<1$--$2~\msun$) are enriched, whereas YMC abundance measurements to date have been performed on red (super)giants with masses $m>2~\msun$. We thus see that the lack of star formation activity in YMCs is an important discriminator between the models, which should be advanced further by observing the chemical abundance patterns of {\it low-mass} stars ($m<1~\msun$) in YMCs.

The additional fact that metal-poor GCs in Fornax, WLM and IKN dwarf galaxies constitute a substantial fraction ($\sim20\%$) of the total metal-poor stellar mass (see \S\ref{sec:gc} and \citealt{larsen12,larsen14}) shows the mass budget problem of the AGB and FRMS models cannot be solved by the loss of $>90\%$ of the initial GC mass by cluster disruption -- otherwise the relics should have become part of the field star population. A solution to the mass budget problem has been proposed by \citet{dercole08} and \citet{dantona13}, who suggested a truncation of the IMF of the enriched population above $m=0.8~\msun$. Such a substantially different IMF has never been observed directly \citep{bastian10}, and the present-day mass functions of GCs are all consistent with a combination of the regular Charbrier-type IMF and dynamical disruption, during which the low-mass stars escape preferentially \citep{baumgardt03,kruijssen09c}. In addition, modifying the IMF to solve the mass budget problem also requires that the enriched stars are not lost by disruption, that the ICMF has a narrow mass range around GC masses, and that the CFE is $\Gamma\sim1$, all of which are inconsistent with observations of nearby YMCs

The EDA model is more consistent with observations of nearby YMCs than the other two models and hence holds up against Occam's Razor more satisfactorily. However, it still has its own set of problems, clearly showing that no definitive model for the observed chemical abundance patterns exists. It is obvious that the evaluation of the above three theories demands explicit predictions that do not only concern chemical abundance patterns, but are also related to nearby YMC formation and the high-redshift galaxy context. Scenarios violating these independent constraints naturally invite criticism. Questions such as how much mass a young GC could possibly accrete, or how much more massive GCs may have been in the past should be answered using both observations of nearby YMCs and the modelling of these processes in the high-pressure context of high-redshift galaxies.

\section{Physical processes governing globular cluster formation and evolution} \label{sec:phys}
A wide range of physical mechanisms has been proposed to be responsible for the formation of GCs, some of which are exclusive to the high-redshift environment, whereas others apply to the local Universe too. In this section, we briefly summarise a selection of formation theories before focussing on the physics relevant for the main hypothesis addressed in this review, namely that GC are the descendants of regular YMC formation at high-redshift.

\subsection{Formation theories} \label{sec:formalt}
As discussed in \S\ref{sec:obs}, the discovery of extremely massive YMCs in galaxy merger remnants gave rise to the idea that GCs may have formed in mergers and could still be forming today. In the context of \S\ref{sec:mtoomre}, this is understood by the increase of the maximum mass-scale for gravitational collapse in the high-pressure environments of galaxy mergers compared to isolated discs, enabling the formation of more massive YMCs. It is now known that mergers are not necessary for high Toomre masses, as these are reached in regular, high-redshift disc galaxies too. Likewise, the formation of GCs was once thought to take place only under special conditions at high-redshift, such as the high Jeans mass following recombination \citep{peebles68}, the compression of gas in strong shocks \citep{gunn80}, or the presence of thermally unstable, metal-poor gas in hot galactic haloes \citep{fall85}. These examples each have their own problems (e.g.~GCs are found in galaxies of too low mass to sustain a hot halo), but they are no longer required to explain the existence of stellar clusters with masses $M>10^6~\msun$.

Another GC formation theory exclusive to the conditions of the high-redshift Universe states that GCs may have formed within their own dark matter haloes before reionization \citep[e.g.][]{peebles84,bromm02,kravtsov05,bekki06,boley09}. This hypothesis may be hard to test, as the dark matter would have been largely removed from the GC due to the \citet{spitzer69} instability between unequal-mass particles (or, put more simply given the large mass difference: the dynamical friction of stars against the dark matter background). However, this model does predict a detectable presence of dark matter in the outskirts of GCs that have always resided in weak tidal fields. Recent observational tests of the stellar kinematics in such GCs rule out the presence of any dark matter reservoir larger than the stellar GC mass \citep{baumgardt09,conroy11b}, ruling out the halo masses $M_{\rm h}\gg M$ necessary to host the formation of GCs.

Finally, the discovery of substantial spreads of heavy element abundances in a small subset of GCs have led to the suggestion that these GCs may be the former nuclei or nuclear clusters of tidally stripped dwarf galaxies \citep[e.g.][]{mackey05,lee09}. A nuclear origin has previously been attributed to the recently-discovered ultra-compact dwarf galaxies \citep[UCDs, see e.g.][]{drinkwater03,misgeld11}. While some studies have advocated the inclusion of UCDs in GC samples \citep[e.g.][]{hilker09,pfeffer14}, UCDs are typically more massive than GCs due to their likely different physical origin -- nuclear clusters are thought to have formed from the merging of GCs at the centre of the galaxy \citep{miocchi06} as well as further gas accretion \citep{hartmann11}, which is consistent with the observed spread in heavy element abundances. Including them may (artificially) increase the GC mass range to higher masses with respect to the Toomre limit that holds for regularly-formed cluster populations (see \S\ref{sec:mtoomre}). None the less, the GC and UCD mass ranges do partially overlap, implying that they can be hard to distinguish at the high-mass end of the GC mass range when chemical indicators are unavailable. The obvious question is what fraction of the total GC population these former nuclear clusters represent. Using indirect arguments, \citet{kruijssen12b} estimate a strict upper limit of $\sim15\%$, but given the small number of GCs with observed heavy element abundance spreads, this number may be substantially lower.

In the remainder of this section, the physics of regular cluster formation and disruption is discussed in the context of GC formation at high redshift, with the eventual intention of constructing a toy model for YMC formation in $z>2$ galaxies, verifying whether these clusters could survive until the present day, and (if so) predicting their present-day properties (see \S\ref{sec:pic}).

\subsection{Regular young massive cluster formation} \label{sec:ymcform}
In brief, theoretical work and numerical simulations of massive cluster formation yield results consistent with the observational picture sketched in \S\ref{sec:ymc}. The current state of cluster formation simulations was recently reviewed in \citet{kruijssen13}, where we concluded that the ingredients required for a representative modelling of cluster formation are turbulent initial conditions, protostellar outflows, radiative feedback from massive stars, and magnetic fields. These have been included to reasonable accuracy in current state-of-the-art simulations of low-mass clusters \citep[e.g.][also see \citealt{tilley07} and \citealt{bonnell08}]{krumholz12c}, but current computational facilities do not yet enable their application to the mass-scales of YMC formation without sacrificing the resolution to resolve the formation of low-mass stars near the hydrogen burning limit \citep[e.g.][]{dale12}. Stellar cluster formation is a multi-scale problem, which obviously requires compromises to be made between the physics included and the dynamic range (and hence the resolution) of the simulation.

Limitations aside, theory and simulations seem to converge to the conclusion that stellar clusters form through the hierarchical fragmentation of the turbulent ISM and the subsequent merging of stellar aggregates on short ($\sim\myr$) time-scales \citep{maclow04,allison09,fellhauer09,maschberger10,fujii12,rogers13,krumholz14}. Gravitationally bound stellar structure arises at the density peaks, due to the combination of sub-virial velocities and high local SFEs \citep{offner09,kruijssen12,girichidis12b}, which enable the stellar groups to remain gravitationally bound upon gas expulsion by feedback \citep{hills80,lada84,goodwin06,baumgardt07,smith13}. In numerical experiments, the initial mass and radius of the bound part of the stellar distribution depends on the implementation of stellar feedback. While the mass spectrum is not expected to depend on the pressure (see below and \citealt{elmegreen97}), there are tentative indications that high-pressure initial conditions lead to more compact clusters than low-pressure conditions, with a rough trend of $R\propto P^{-0.25}$ \citep{zubovas14}. However, a more extensive parameter survey is necessary to provide a definitive answer.

There exist two analytical models aimed at characterising cluster formation on galactic scales. \citet{hopkins13} employs the excursion-set formalism that has been applied to cosmic structure for decades \citep{press74} to address the spatial clustering of protostellar cores as it arises from the turbulent ISM. He reproduces the observed auto and cross-correlation functions seen in observed star-forming regions. Using this formalism, it is found that only a small fraction ($10\%$) of all star formation occurs in isolation. Whereas this work clearly shows that spatial clustering is a fundamental property of the star formation process in large-scale driven, turbulent clouds, it does not make any statements regarding the fraction of star formation that occurs in {\it gravitationally bound} clusters (the aforementioned CFE). When aiming to understand the formation and long-term survival of GCs, it is necessary to obtain a theoretical model for the CFE. Such a model was presented in \citet{kruijssen12d}, where we used the volume density probability distribution function (PDF) of the ISM in equilibrium galaxy discs to predict at which densities bound stellar cluster formation may take place. The critical question is above which densities the free-fall time is short enough to achieve a high local SFE (and hence a high bound fraction of stars) before feedback halts star formation. Integration over the full density PDF then yields the fraction of star formation occurring in bound clusters. This model predicts a CFE that increases with the gas pressure (or surface density), in quantitative agreement with both earlier and subsequent observational studies (see \S\ref{sec:cfe}). While this gives some confidence that the model includes some of the relevant physics, the obvious shortcoming is that without additional information it is unable to predict spatial clustering, cluster masses, and cluster radii due to being based on one-point statistics (i.e.~the density PDF). As such, the \citet{hopkins13} and \citet{kruijssen12d} models describe fully complementary aspects of cluster formation -- the formulation of a combined model should be a priority in the near future \citep[as was recently proposed by][]{krumholz14}.

The physics underpinning the ICMF is theoretically reasonably well-understood. It was shown by \citet{elmegreen96} that the $\alpha=-2$ power law arises naturally from the fractal structure of the ISM from which stellar clusters form \citep[e.g.~by the hierarchical merging of smaller structure, also see][]{fujii13}. Their analysis was extended by \citet{fall10}, who considered the translation from the GMC mass function to the ICMF by accounting for the mass dependences of the cloud lifetime and the SFE. These authors show that the $\alpha=-1.7$ slope of the GMC mass function is converted into $\alpha=-2$ for stellar clusters if (1) momentum-driven feedback is responsible for halting star formation (for energy-driven feedback $\alpha=-1.8$ is obtained) and (2) star-forming regions have a constant surface density (i.e.~$M\propto R^{1/2}$). The former is true for cluster masses $M\geq{\rm few}\times10^4~\msun$ and gas densities $\Sigma\geq{\rm few}\times10^2~\msun~\pc^{-2}$, consistent with the likely progenitor masses and formation environments of GCs. However, while the latter requirement ($M\propto R^{1/2}$) is certainly true for the GMCs from which stars form, it is unclear if it holds for the bound clusters emerging from the cluster formation process, for which no mass-radius relation is found (see \S\ref{sec:rad}).

No theoretical model exists for the maximum cluster mass or exponential truncation mass of the ICMF at the high-mass end. However, if the above mapping of the GMC mass function to the ICMF is reasonable, a similar argument can be made for obtaining the maximum cluster mass from the two-dimensional Jeans mass or Toomre mass in galaxies, i.e.~the largest mass-scale to overcome shear and become self-gravitating. As was shown in \S\ref{sec:mtoomre} and Figure~\ref{fig:mtoomre}, this translation seems consistent with observations, suggesting that the maximum cluster mass is related to the Toomre mass according to equation~(\ref{eq:mmax}). Given a theory for the CFE and assuming that the SFE is roughly constant (see \S\ref{sec:mtoomre}), the remaining question is whether the Toomre mass is indeed a physically relevant quantity despite the simple analytical derivation. Using high-resolution magnetohydrodynamic simulations, \citet{kim01} find that the Toomre argument applies to magnetised discs, although the extent depends on the strengths of the magnetic field and shear. The practical advantage of the Toomre mass as expressed in equation~(\ref{eq:mtoomre}) is that is can easily be written in terms of observable parameters such as the gas surface density, angular velocity, and the Toomre $Q$ disc stability parameter \citep{krumholz05}, which is close to unity for galaxy discs in hydrostatic equilibrium. As a result, it not only provides a physically meaningful, but also a versatile reference mass for models of GMC and stellar cluster formation.

\subsection{Cluster disruption in gas-rich, high-pressure environments} \label{sec:gasrich}
In high-pressure, gas rich environments, cluster disruption (and survival) is dominated by transient tidal perturbations (often referred to as `impulsive' or `tidal' shocks) due to encounters with GMCs (or, on a smaller scale, the density peaks within GMCs). There exists a large body of literature on cluster disruption by tidal shocks \citep[e.g.][]{spitzer58,ostriker72,weinberg94b,kundic95,gnedin97,gieles06,gieles07}, which provides a quantitatively consistent formalism for modelling cluster disruption by tidal shocks. The resulting disruption time-scale scales linearly with the cluster density, implying that models describing the evolution of cluster masses only need to assume some relation between the cluster mass and radius.

Disruption by tidal shocks is particularly relevant for the early phases of GC evolution, when they are still associated with the high-pressure environments in which they formed and therefore should be expected to undergo rapid disruption \citep[e.g.][]{elmegreen10}, in strong contrast with their later evolution in gas-poor galaxy haloes. Because stellar clusters (and YMCs in particular) naturally form under high-pressure conditions, their migration away from their birth environments in general leads to a decrease of the cluster disruption rate with cluster age \citep{elmegreen10b,kruijssen11}. This process has been referred to as the `cruel cradle effect', reflecting that the disruption rate of a cluster population must peak at young ages -- clusters that after some time have escaped their birth environments (e.g.~into gas-poor galaxy haloes) experience less disruption by tidal shocks and may even be subject to disruption by evaporation only (see \S\ref{sec:gaspoor} below), leading to strongly increased lifetimes. Of course, if these clusters can survive as long as a Hubble time they may represent the GCs that are observed today. Cluster migration therefore plays a central role in the problem of GC formation.

Another source of YMC disruption in the compact, high-pressure discs of high-redshift galaxies is the dynamical friction of massive clusters against the background matter \citep[e.g.][]{tremaine75,goerdt06}. If the clusters survive the resulting inspiral all the way to the host galaxy's centre without being disrupted by tidal shocking or stripping, dynamical friction may contribute substantially to the growth of nuclear clusters often observed in galaxy centres \citep[e.g.][]{miocchi06,capuzzodolcetta08,antonini13}. Given a circular velocity $V$, the initial galactocentric radius $R$ out to which GCs may have spiralled in to the galactic centre within a time-scale $t$ is roughly given by
\begin{equation}
\label{eq:rdf}
R=0.9~\kpc\left(\frac{t}{2~\gyr}\right)^{1/2}\left(\frac{V}{100~\kms}\right)^{-1/2}\left(\frac{M}{10^6~\msun}\right)^{1/2} ,
\end{equation}
showing that dynamical friction may have weakly affected the high-mass end of the cluster mass range in compact $R<2~\kpc$ galaxies, but did not contribute substantially to the disruption of young GCs unless (1) the migration time-scale $t_{\rm mig}\gg2~\gyr$ or (2) the host galaxy's circular velocity $V\ll100~\kms$ (e.g.~in dwarf galaxies).

\subsection{Cluster migration} \label{sec:mig}
The rapid disruption of YMCs (or young GCs) in their gas-rich, high-pressure birth environment comes to an end when they migrate into a less violent environment (e.g.~gas-poor galaxy haloes). Note that the use of the term `migration' strictly refers to escaping the part of phase space that is occupied by the gas. When this happens, the long-term survival chances of a YMC improve substantially, as it no longer coincides in position-velocity space with the gas-rich, star-forming (hence disruptive) component of the host galaxy. Numerical simulations investigating the formation and/or disruption and survival of stellar clusters during hierarchical galaxy formation \citep{kravtsov05,prieto08,rieder13} or galaxy mergers \citep{li04,kruijssen12c,renaud13} reveal the efficient migration of pre-existing clusters during mergers with mass ratios $x\equiv M_{\rm host}/M_{\rm other}<3$ (where $x<1$ indicates that the host galaxy is merging with a more massive galaxy). Galaxies undergoing frequent galaxy mergers host cluster populations characterised by lower disruption rates than those in more quiescent environments \citep{rieder13}, which is an immediate result of cluster migration into the tidal debris in the galaxy halo \citep[also see][]{renaud13}.

\begin{figure}
\center\includegraphics[width=11.4cm]{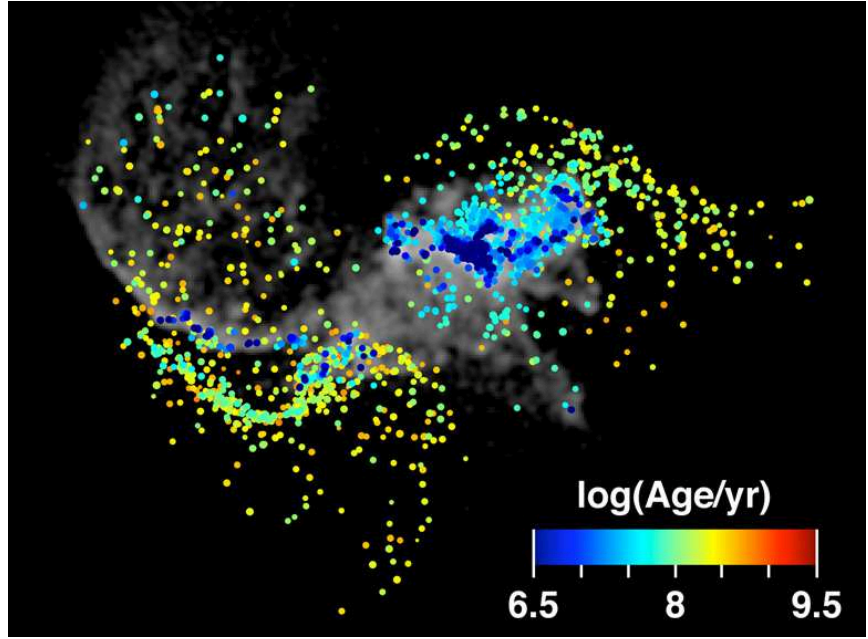}
\caption{This figure shows how stellar clusters migrate into the host galaxy halo during a galaxy merger. Shown is a snapshot from an equal-mass, Milky Way-like galaxy merger simulation that includes a sub-grid model for the formation and disruption of the cluster population, covering the entire cluster mass range of $M\sim10^2$--$10^6~\msun$. The time of the snapshot corresponds to the first pericentre passage of the two galaxies. Coloured dots indicate stellar clusters, colour-coded by their ages as indicated by the legend. The grey-shaded material indicates the gas. Note how intermediate-age clusters that were formed before the merger escape into the quiet waters of the halo, whereas young clusters reside in the disruptive, high-pressure environment of the galaxy discs. Taken from \citet[Figure~1]{kruijssen12c}, which includes the full time sequence of this merger model as well as a movie.} \label{fig:mig}
\end{figure}
Whereas the above two simulations do not include a gas component and are therefore restricted to studying cluster disruption outside of their formation environment, they are complemented by the simulations presented in \citet{kruijssen12c}, where we addressed mainly the early evolution of the cluster population in galaxy mergers. It was found that the destruction rate peaks strongly briefly after the pericentre passages of the merging galaxies, as well as during their coalescence. This is caused by the rapid inflow of gas towards the galaxy centres and the corresponding increase of the gas density and the SFR, which produces large numbers of clusters but also leads to their destruction on short time-scales due to the tidal shocking by encounters with dense gas. As a result, the clusters surviving the merger are those that formed before the peak of the interaction \citep[also see Figure~18 of][where this was shown explicitly]{kruijssen11}, which then migrate into the halo during the pericentre passages and final coalescence (see Figure~\ref{fig:mig}). The different numerical experiments discussed here thus give a consistent result -- the survival chances of {\it pre-existing} clusters are increased by their migration occurring during galaxy mergers. This provides an interesting contrast with the earlier view that the YMC progenitors of present-day GCs formed in mergers \citep{ashman92} -- in the new picture, mergers do not necessarily {\it form} these YMCs, but {\it liberate} them into the halo so that they survive and become GCs.\footnote{The reason that the most extreme YMCs at $z=0$ are generally observed in galaxy mergers is that these are the systems with the highest mass-scales for gravitational collapse in the nearby Universe (see \S\ref{sec:mtoomre}). Such conditions were normal in galaxies at $z>2$, enabling the formation of GC-like YMCs in normal disc galaxies across most of the mass and morphology range. This may solve the problem that the spatial distribution of the metal-rich GC population in the Milky Way cannot be reproduced if metal-rich GC formation only takes place in galaxy mergers \citep{griffen10}.}

Galaxy mergers occur frequently in a $\Lambda$CDM Universe and therefore they provide a natural mechanism to end the initial, rapid-disruption phase after cluster formation. How effective this mechanism is exactly depends on the precise merger time-scale. Depending on the relevant subset of mergers considered (mass ratio, redshift range, host galaxy mass, etc.), the merger rate varies from $\nu_{\rm merge}=10^{-1}$--$10^1~\gyr^{-1}$ \citep[e.g.][]{genel09}. The corresponding merger time-scale is long ($t_{\rm merge}\equiv\nu_{\rm merge}^{-1}=10^8$--$10^{10}~\yr$), implying that a more detailed calculation is necessary to establish whether YMCs can survive long enough to be `liberated' into the host galaxy halo by mergers. Qualitatively though, the merger time-scale increases with cosmic time and with galaxy mass \citep{genel09,fakhouri10}, indicating that GC formation (i.e.~the migration of surviving YMCs into low-disruption environments) was most efficient in high redshift, low-mass galaxies.

Next to ensuring the long-term survival of massive clusters, the hierarchical growth of galaxies also plays a key role in determining the properties of present-day GC systems. \citet{searle78} were the first to propose that the present day (metal-poor) GC populations of galaxies like the Milky Way may be constituted by sub-populations having formed in different progenitor galaxies. The idea that these progenitors simultaneously contributed to both the hierarchical growth of the GC system and the stellar halo has since been confirmed by observational and numerical work \citep[e.g.][]{forbes01,muratov10,forbes10,tonini13,li14,katz14}. This new scenario, in which the metal-poor GCs were accreted from cannibalised, low-mass dwarf galaxies, supplements the longer-standing picture in which the metal-rich part of the GC population is associated with the host galaxy's spheroid \citep[e.g.][]{forbes01}. As discussed in \S\ref{sec:feh}, these new models describing the hierarchical growth of GC systems and their corresponding migration automatically reproduce the observed bimodality of GC colours and metallicities. However, these models do not contain a description for GC formation itself, nor do they describe the initial, rapid-disruption phase of YMC evolution. They must therefore assume some relation between the host galaxy mass and the initialisation of the high-redshift GC population in the host galaxy halo. Given the current understanding of massive cluster formation, it is inevitable that GC formation models will improve substantially on these assumptions in the near future.

\subsection{Cluster disruption in gas-poor environments} \label{sec:gaspoor}
After having migrated into the host galaxy halo, GCs dissolve at a modest rate due to combination of evaporation, disc shocks and bulge shocks \citep{spitzer87,gnedin97,gnedin99b,fall01}, where the latter two terms indicate specific subsets of tidal shocks in which the transient perturbation is caused by crossing the host galaxy's disc or passing through the minimum gravitational potential of the GC orbit at pericentre, respectively. Multiple studies have shown that the mass loss rate due to disc shocking is negligible compared to the mass loss by evaporation \citep[e.g.][]{dinescu99}, whereas bulge shocking only dominates over evaporation if the eccentricity of the GC orbit is $e>0.5$ \citep[e.g.][]{baumgardt03}.

A wide range of models assumes that GC evaporation occurs on a two-body relaxation time \citep[e.g.][]{chernoff90,gnedin97,fall01,mclaughlin08,prieto08,muratov10,li14}, i.e.
\begin{equation}
\label{eq:trh}
t_{\rm dis}\propto t_{\rm rh}\propto \frac{N^{1/2}}{\ln{\Lambda}}r_{\rm h}^{3/2}\propto \frac{N}{\ln{\Lambda}}\rho_{\rm h}^{-1/2}\propto \frac{N}{\ln{\Lambda}}t_{\rm cr} ,
\end{equation}
where $N$ is the number of stars, $r_{\rm h}$ is the cluster half-mass radius, $\rho_{\rm h}$ is the half-mass density, $\ln{\Lambda}$ (with $\Lambda\propto N$) is the Coulomb logarithm, and $t_{\rm cr}$ is the crossing time at $r_{\rm h}$. If indeed $t_{\rm dis}\propto t_{\rm rh}$, then high-density, low-mass clusters have the shortest lifetimes, independent of the properties of the host galaxy. The aforementioned models have used the environmental independence of the GC disruption time implied by equation~(\ref{eq:trh}) to describe the evaporation of low-mass clusters at a constant, universal rate ${\rm d}M/{\rm d}t\equiv -M/t_{\rm dis}\propto\rho_{\rm h}^{1/2}$ for a Hubble time of dynamical evolution, thereby evolving the power-law ICMF into a peaked GCMF with a universal peak mass.

However, the above picture is too simplified for a variety of reasons. \citet{baumgardt01} showed that stars with energies larger than the escape energy require a finite number of crossing times to reach one of the Lagrangian points and escape \citep{fukushige00}, leading to a disruption time scaling as
\begin{equation}
\label{eq:tdis}
t_{\rm dis}\propto t_{\rm rh}^{3/4}t_{\rm cr}^{1/4}\propto\left(\frac{N}{\ln{\Lambda}}\right)^{3/4}t_{\rm cr}\rightarrow M^{0.6{\rm -}0.7}\Omega^{-1} ,
\end{equation}
where $\Omega$ is the angular velocity of the orbit ($\Omega\equiv V/R$) and the final term comes from comparisons to $N$-body simulations, initially assuming that the half-mass radius scales with the Jacobi radius $r_{\rm J}$ \citep{lamers05}, but later on clusters with a broad range of $r_{\rm h}/r_{\rm J}$ ratios were included \citep{lamers10}. The insensitivity of the disruption time-scale to $r_{\rm h}/r_{\rm J}$ arises because the radius dependence of $t_{\rm rh}\propto r_{\rm h}^{3/2}$ is cancelled by an identical scaling of the number of escapers per unit $t_{\rm rh}$, yielding a constant escape rate per unit absolute time \citep{gieles08}. The scaling of equation~(\ref{eq:tdis}) has been confirmed by $N$-body simulations of clusters dissolving on a variety of stellar mass functions, stellar evolution recipes, and (eccentric and circular) orbits in galactic background potentials. Perhaps most importantly, these simulations have shown that the proportionality constants of equations~(\ref{eq:trh}) and~(\ref{eq:tdis}) depend on the orbital angular velocity $\Omega$, whereas the mass loss rate due to bulge shocking introduces an additional dependence on the orbital eccentricity $e$. GC evaporation therefore depends on the galactic environment and cannot be universal \citep[e.g.][]{vesperini97b,portegieszwart98,baumgardt03,madrid14}. Consequently, GC evaporation also cannot explain the universality of the GC characteristic mass-scale -- this is consistent with the commonly found result that the present-day mass loss rate of GCs in galaxy haloes is too low to evolve an initial YMC population into a GC population over the course of a Hubble time, unless the orbits are much more eccentric than is actually observed \citep[also see \citealt{mieske14}]{fall01,vesperini03,kruijssen09b}. If the present-day GC population is the product of regular YMC formation at high redshift, an early period of enhanced cluster disruption (\S\ref{sec:gasrich}) is therefore a necessity.

Is the evolution of GCs in the silent waters of galaxy haloes important at all? For GC masses this may be an open question, but the late dynamical evolution of GCs certainly plays a central role in their structural evolution. $N$-body simulations and theory show that after an initial phase characterised by expansion, the half-mass radii of GCs become roughly proportional to their tidal radii, which are set by the haloes in which they presently reside \citep{gieles11b}. In this tidally-limited phase, the orbital parameters (e.g.~the eccentricity), the background potential, and the position along the orbit all affect the present-day GC radius \citep{madrid12}. Not all GCs have reached this final phase yet \citep{baumgardt10}, indicating that the structural evolution of GCs in the context of their host galaxy haloes is very much ongoing at the present cosmic epoch.

\section{Shaken, then stirred: a two-phase scenario for globular cluster formation} \label{sec:pic}
In this section, we condense the observational (\S\ref{sec:obs}) and theoretical (\S\ref{sec:phys}) constraints on GC formation into a coherent picture of GC formation. In a companion paper \citep{kruijssen14c}, we quantify the proposed scenario and show that it reproduces the observed properties of the GC population at $z=0$.
\begin{itemize}
\item[(i)]
The majority of the GC progenitor clusters are formed in the high-pressure, gas-rich discs of `normal', star-forming galaxies at $z>2$, in which each cluster-forming event occurs by the hierarchical merging of smaller structures on a time-scale of a few~$\myr$ (see \S\ref{sec:hierarchy} and \S\ref{sec:ymctime}). The high gas pressures and densities lead to a high CFE as well as a high Toomre mass, enabling the formation of stellar clusters with initial masses $M_{\rm i}\gtrsim10^6~\msun$ at the high-mass end of a power-law ICMF with index $\alpha=-2$ (see \S\ref{sec:cfe}--\ref{sec:mtoomre} and \S\ref{sec:ymcform}). Using the galaxy mass-metallicity relation at $z\sim3$ (see \S\ref{sec:hiz}), we find that these clusters have metallicities between $\feh=-2.5$ and $\feh=-0.5$ for galaxies with stellar masses in the range $M_{\rm star}=10^7$--$10^{11}~\msun$ (see \S\ref{sec:sites}).
\item[(ii)]
At a maximum cluster mass of $M\gtrsim10^6~\msun$, the Toomre mass should be $M_{\rm T}\gtrsim10^{7.5}~\msun$ and the gas pressure $P/k\sim10^7~{\rm K}~\cmc$ (see \S\ref{sec:conditions}). For a gas disc in hydrostatic equilibrium with a typical angular velocity of $\Omega\sim0.1~\myr^{-1}$ \citep{forsterschreiber09}, this corresponds to a surface density of $\Sigma\sim300~\msun~\pc^{-2}$ and a mid-plane volume density of $\rho_{\rm ISM}\sim2~\msun~\pc^{-3}$. These densities are over an order of magnitude higher than in the disc of the Milky Way, but represent common conditions in high-redshift disc galaxies (see \S\ref{sec:sites}). Because the maximum GC mass increases with the galaxy mass and the mean metallicity of the GC population (see \S\ref{sec:gcmf} and \S\ref{sec:feh}), the gas pressure during GC formation likely increased with the host galaxy mass and metallicity too. As a result, more massive and more metal-rich galaxies may have produced a larger fraction of their star formation in the form of bound clusters (cf.~\S\ref{sec:cfez}), with a higher maximum cluster mass (cf.~\S\ref{sec:mtoomrez}).
\item[(iii)]
After their formation, the stellar clusters reside in the gas-rich, high-pressure disc of their host galaxy. This episode (Phase~1) is characterised by rapid cluster disruption, due to strong tidal shocks experienced during the frequent encounters with gas density peaks (see \S\ref{sec:hizdis} and \S\ref{sec:gasrich}). In the absence of a strong relation between the cluster mass and radius (see \S\ref{sec:rad}), the disruption time-scale increases near-linearly with the cluster mass, indicating that low-mass clusters are rapidly destroyed whereas the most massive clusters survive. The rate of cluster disruption increases with the ambient gas density, and hence with the host galaxy mass and metallicity, leading to a lower survival fraction in more massive and metal-rich galaxies (cf.~\S\ref{sec:tn}).
\item[(iv)]
The rapid disruption phase continues until the clusters migrate into the galaxy halo due to a galaxy merger with another galaxy at least as massive as the host galaxy. For very dissimilar galaxy masses, the clusters are tidally stripped from their host and enter the halo of the more massive galaxy. If the galaxy mass ratio is close to unity, the clusters are ejected into the halo by violent relaxation. Because the merger rate with equal-mass or more massive galaxies decreases with the galaxy mass, clusters formed in massive, metal-rich galaxies experience a longer rapid-disruption phase than those formed in low-mass, metal-poor galaxies (see \S\ref{sec:mig}). Again, this leads to a lower survival fraction in more massive and metal-rich galaxies (cf.~\S\ref{sec:tn}). The stochastic nature of galaxy mergers could introduce some galaxy-to-galaxy variation of the GC survival fraction, especially in low-mass (and hence low-$N_{\rm GC}$ galaxies. This may provide a possible explanation for the high specific frequency seen in some dwarf galaxies (see \S\ref{sec:tn}).
\item[(v)]
After the clusters have migrated into the galaxy halo, they are no longer disrupted by strong tidal shocks, which greatly increases their survival chances. These new `GCs' remain in the halo until the present day (Phase~2), during an episode characterised by slow disruption due to evaporation (see \S\ref{sec:gaspoor}). Despite the fact that GCs spend a long time in the host galaxy halo, the disruption rate integrated over this phase is small (see \S\ref{sec:tn}). As a result, the cluster masses evolve only by a minor amount, whereas their structural evolution can proceed freely due to the scarcity of transient tidal perturbations.
\item[(vi)]
The slow disruption in the galaxy halo implies that the near-universal peak mass of the GCMF is an imprint of the rapid-disruption phase in the host galaxy disc, when low-mass clusters were efficiently destroyed. Because the formation environments of GCs were more similar than the wide range of tidal fields they presently reside in, this could explain why the peak mass is near-universal. The GC populations of galaxies may continue to grow by the accretion of satellite galaxies. Statistically, minor mergers are more likely to occur (see \S\ref{sec:mig}), implying that accreted GCs are typically more metal-poor than the GC population native to the host galaxy. As stated in point~(iv), these metal-poor galaxies should have high cluster survival fractions (i.e.~high specific frequencies). This naturally leads to a bimodality in GC colour and metallicity (see \S\ref{sec:feh}).
\item[(vii)]
Finally, the two-phase picture of GC formation presented here suggests that GC formation was more efficient at high redshift due to a combination of (1) consistently high Toomre masses, which enabled the formation of massive, long-lived stellar clusters, and (2) high galaxy merger rates, which facilitated the rapid migration of clusters into the host galaxy halo, thus allowing them to survive for a Hubble time. However, it also implies that GC formation can still take place in the present-day Universe, although it is much rarer due to the low Toomre masses and galaxy merger rates compared to those of $z>2$ galaxies.
\end{itemize}

Recent reviews by \citet{brodie06} and \citet{harris10} made significant, mainly observational steps towards understanding GC formation in the context of galaxy formation and evolution. In particular, \citet{harris10} lists several open physical questions, relating to e.g.~the origin of metallicity bimodality and the decrease of the specific frequency with metallicity, as well as expressing the need for a holistic framework for GC formation and evolution from birth to death. While the above scenario for GC formation is entirely qualitative and follows broad brush strokes, it does unite the present observational and theoretical constraints on the cosmological assembly of GC populations from the three directions discussed in this review: YMC formation in nearby galaxies, star formation in high-redshift galaxies, and the present-day properties of GC populations. At the same time, it addresses several of the questions posited by \citet{harris10}. It is an inviting story because Occam's Razor is satisfied -- no new physics were introduced to construct this scenario. The obvious question is if it works in practice. This question is addressed in \citet{kruijssen14c}, where a quantitative model of the above scenario is presented, enabling a comparison to observations of nearby GC populations as well as the formulation of testable predictions for future work.

The presently available evidence suggests that GC populations may well arise naturally from a combination of the observed regular star formation activity at $z>2$ and the hierarchical growth of galaxies. As discussed at length in \S\ref{sec:obs} and \S\ref{sec:phys}, it has been pointed out before that some properties of the present-day GC population may emerge naturally from `normal' star and galaxy formation physics \citep[e.g.][]{tonini13,li14,katz14}. These models emphasise the importance of hierarchical galaxy growth for explaining the properties of the GC population in a similar way as is done in our two-phase scenario for GC formation. However, none of them contains a physical description of the formation of the GC progenitor clusters, implying these models effectively start at point~(v) of the above scenario. To circumvent this, they have to assume the $z=0$ relation between the specific frequency and the galaxy mass to initialise the GC population and/or omit the environmental variation of cluster disruption. A quantitative formulation of the two-phase model would not require these assumptions to be made.

Testing the hypothesis that GCs are the natural outcomes of the galaxy formation process should be a prime target for future work. In addition, several other open questions remain. How do the multiple stellar populations observed in GCs arise? Through which other processes than regular YMC formation may GC-like objects form? We have seen in \S\ref{sec:chem} that some GCs (such as $\omega$Cen) display heavy-element abundance variations, suggesting that they are the former nuclear clusters of dwarf galaxies. Even though these GCs seem to represent a minority, they provide a case in point that the present-day GC population may have resulted from more than a single physical process. Charting the relative contributions of these GC formation mechanisms to the population at $z=0$ should be a priority.

Above all, it is clear that the process of GC formation covers a broad range of spatial scales and physical mechanisms. The physics of GC formation will not be unveiled by isolating problems and solving them without context. GCs did not form in isolation, nor can the physics of their formation be derived from the properties of their host galaxies alone. Instead, the most promising way forward is to keep combining the rich variety of methods and approaches, both by drawing from those available in the literature today and by developing new avenues.

\ack
JMDK is greatly indebted to Nate Bastian, Jim Dale, Bruce Elmegreen, Mark Gieles, Henny Lamers, Chervin Laporte, S\o ren Larsen, Steve Longmore, Simon Portegies Zwart, and Marina Rejkuba for several years of discussions from which large parts of this work were drawn. Both referees, as well as Nate Bastian, Steve Longmore, Fran\c cois Schweizer and Enrico Vesperini are gratefully acknowledged for their careful reading of the manuscript. Gabriella De Lucia, Andr\'{e}s Jord\'{a}n, and Eric Peng are thanked for granting permission for the reuse of their figures as Figure~\ref{fig:tree},~\ref{fig:mpeak}, and~\ref{fig:bimodality} in this work. JMDK thanks Shy Genel, Michael Hilker, Phil Hopkins, Mark Krumholz, Claudia Lagos, Ben Moster, Thorsten Naab, Jay Strader, Linda Tacconi, and Simon White for helpful conversations and suggestions.

\bibliographystyle{mn2e}
\bibliography{mybib}

\end{document}